\documentclass[amsmath,amssymb,aps,floats,amsfonts,notitlepage,superscriptaddress,eqsecnum,nofootinbib]{revtex4-1}
\usepackage{graphicx}       % Include figure files
\usepackage{dcolumn}        % Align table columns on decimal point
\usepackage{bm}             % bold math
\usepackage{amssymb}
\usepackage{amstext}
\usepackage{tensor}
\usepackage[colorlinks]{hyperref}
\usepackage{color}
\usepackage[usenames,dvipsnames,svgnames,table]{xcolor}
\usepackage{float}
\allowdisplaybreaks[1]

\definecolor{CiteColor}{rgb}{0,0.5,0}
\hypersetup{citecolor=CiteColor}
\definecolor{RefColor}{rgb}{0.55,0,0}
\hypersetup{linkcolor=RefColor}
\definecolor{darkgreen}{rgb}{0.2,0.7,0.2}

\newcommand{\alp}{\alpha}

\newcommand{\eps}{\epsilon}

\newcommand{\en}{\mathcal{E}_0}
\newcommand{\ang}{\mathcal{L}_0}

\newcommand{\sdiff}[2]  {\frac{d^2 #1}{d #2^2}}

\newcommand{\beq}{\begin{equation}}
\newcommand{\eeq}{\end{equation}}

\newcommand{\EE}{\mathcal{E}}
\newcommand{\BB}{\mathcal{B}}
\newcommand{\XX}{\chi}
\newcommand{\RR}{\bar{\mathcal{R}}}

\def\etal{\textit{et al.}}

\begin{document}

\title{Octupolar invariants for compact binaries on quasi-circular orbits}

\author{Patrick Nolan}
\affiliation{School of Mathematical Sciences and Complex \& Adaptive Systems Laboratory, University College Dublin, Belfield, Dublin 4, Ireland.}

\author{Chris Kavanagh}
\affiliation{School of Mathematical Sciences and Complex \& Adaptive Systems Laboratory, University College Dublin, Belfield, Dublin 4, Ireland.}

\author{Sam R.~Dolan}
\affiliation{Consortium for Fundamental Physics, School of Mathematics and Statistics,
University of Sheffield, Hicks Building, Hounsfield Road, Sheffield S3 7RH, United Kingdom.}

\author{Adrian C.~Ottewill}
\affiliation{School of Mathematical Sciences and Complex \& Adaptive Systems Laboratory, University College Dublin, Belfield, Dublin 4, Ireland.}

\author{Niels Warburton}
\affiliation{MIT Kavli Institute for Astrophysics and Space Research, Massachusetts Institute of Technology, Cambridge, MA 02139, USA}
\affiliation{School of Mathematical Sciences and Complex \& Adaptive Systems Laboratory, University College Dublin, Belfield, Dublin 4, Ireland.}

\author{Barry Wardell}
\affiliation{School of Mathematical Sciences and Complex \& Adaptive Systems Laboratory, University College Dublin, Belfield, Dublin 4, Ireland.}
\affiliation{Department of Astronomy, Cornell University, Ithaca, NY 14853, USA.}

\date{\today}

\begin{abstract}
We extend the gravitational self-force methodology to identify and compute new $\mathcal{O}(\mu)$ tidal invariants for a compact body of mass $\mu$ on a quasi-circular orbit about a black hole of mass $M \gg \mu$. In the octupolar sector we find seven new degrees of freedom, made up of 3+3 conservative/dissipative `electric' invariants and 3+1 `magnetic' invariants, satisfying 1+1 and 1+0 trace conditions. After formulating for equatorial circular orbits on Kerr spacetime, we calculate explicitly for Schwarzschild spacetime. We employ both Lorenz gauge and Regge-Wheeler gauge numerical codes, and the functional series method of Mano, Suzuki and Takasugi. We present (i) highly-accurate numerical data and (ii) high-order analytical post-Newtonian expansions. We demonstrate consistency between numerical and analytic results, and prior work. We explore the application of these invariants in effective one-body models, and binary black hole initial-data formulations, and conclude with a discussion of future work.
\end{abstract}

\maketitle

\section{Introduction}

The prospect of `first light' at gravitational wave detectors has spurred much work on the gravitational two-body problem in relativity. It is now a decade since the first (complete) simulations of binary black hole (BH) inspirals and mergers in numerical relativity (NR) \cite{Pretorius:2005gq}. Such simulations have revealed strong-field phenomenology, such as `superkicks' \cite{Brugmann:2007zj}, and have provided template gravitational waveforms. Yet, it may be argued, numerical relativity has also highlighted the `unreasonable effectiveness' of both post-Newtonian (PN) theory \cite{Will:2011nz}, and the Effective One-Body (EOB) model \cite{Hinderer:2013uwa}. 

BH-BH binaries, and their waveforms, are described by parameters including the masses $M$, $\mu$, spins, orbital parameters ($p$, $e$), etc. The parameter space expands for BH-neutron star (NS) binaries -- a key target for detection in 2016 \cite{TheLIGOScientific:2014jea} -- as tidal interactions also play an important role \cite{Bernuzzi:2014owa, Landry:2015cva}. Semi-analytic models, such as the EOB model, allow for much finer-grained coverage of parameter space than would be possible with (computationally-expensive) NR simulations alone. In addition, effective models can bring physical insight \cite{Schmidt:2012rh, Hannam:2013oca, Schmidt:2014iyl}. For \emph{real-time} data analysis it may be necessary to blend effective models with surrogate/emulator models \cite{Cole:2014qha, Blackman:2015pia} and careful analysis of modelling uncertainties \cite{Moore:2014pda}. 

By design, the EOB model \cite{Buonanno:1998gg, Damour:2012ky, Taracchini:2013rva, Damour:2013hea, Damour:2015isa} incorporates under-determined functional relationships, which are `calibrated' with PN expansions and numerical data. Recently, it was shown that invariant quantities computed via the Gravitational Self-Force (GSF) methodology \cite{Poisson:2011nh, Barack:2009ux, Thornburg:2011qk} can be used for exactly this purpose \cite{Damour:2009sm, Barack:2010ny, Akcay:2012ea, Damour:2012ky, Bini:2014ica, Bini:2014zxa}. In fact, as the GSF methodology is designed to provide highly-accurate strong-field data in the extreme mass-ratio regime \cite{Mino:1996nk, Quinn:1996am}, it provides complementary constraints to PN and NR approaches, which excel in the weak-field and comparable mass-ratio regimes, respectively \cite{Tiec:2014lba}. Thus, new GSF data, nominally limited in scope to the extreme-mass ratio regime, $\mu/M \ll 1$, may immediately be applied to enhance models of comparable-mass inspirals, required for data analysis at, e.g., Advanced LIGO \cite{TheLIGOScientific:2014jea}. 

In recent years, a growing number of invariant quantities, associated with geodesic orbits in black hole spacetimes perturbed through linear order $\mathcal{O}(\mu / M)$, have been extracted from GSF theory. For quasi-circular orbits on Schwarzschild, these include (i) the redshift invariant \cite{Detweiler:2008ft, Sago:2008id}, (ii) the shift in the innermost stable circular orbit \cite{Barack:2009ey}, (iii) the periastron advance (of a mildly-eccentric orbit) \cite{Barack:2009ey, Barack:2011ed}, (iv) the geodetic spin-precession invariant \cite{Dolan:2013roa, Bini:2014ica, Bini:2015mza}, (v) tidal eigenvalues \cite{Dolan:2014pja, Bini:2014zxa, Bini:2015kja}, (vi) certain octupolar invariants \cite{Bini:2014zxa, Bini:2015kja}. Recently, (i) has been computed for eccentric orbits \cite{Barack:2011ed, Akcay:2015pza}, and (i)--(ii) have been computed for equatorial quasi-circular orbits on Kerr spacetime \cite{Isoyama:2014mja}. 

In 2008, the GSF redshift invariant at $\mathcal{O}(\mu/M)$ was compared against a post-Newtonian series at 3PN order (i.e.,~$\mathcal{O}(v^6/c^6)$) \cite{Detweiler:2008ft}. Many further PN expansions have followed for invariants (i)--(vi) at very high PN orders \cite{Johnson-McDaniel:2015vva,
Shah:2015nva,Kavanagh:2015lva,Bini:2015mza, Dolan:2014pja,
Dolan:2013roa, Bini:2014zxa, Bini:2014ica}. An `arms race' between numerical (GSF) and analytical (PN) approaches has developed, enabling precise comparisons of high-order coefficients \cite{Johnson-McDaniel:2015vva, Kavanagh:2015lva, Bini:2015mza, Shah:2015nva, Dolan:2014pja}. Such comparisons are invaluable in quality assurance, as they have been used to correct small errors in both GSF calculations \cite{Dolan:2014pja} and PN expansions \cite{Bini:2015mza}. Furthermore, in the `experimental mathematics' approach \cite{Johnson-McDaniel:2015vva, Bailey:2010aj}, high-order PN coefficients may be extracted in closed (transcendental) form from exquisitely-precise numerical GSF calculations.

The purpose of this paper is to classify and compute GSF invariants at `octupolar' order, i.e., featuring three derivatives of the metric, or equivalently, first derivatives of the Riemann tensor. This sector has been previously considered by Johnson-McDaniel \emph{et al.} \cite{JohnsonMcDaniel:2009dq} and Bini \& Damour \cite{Bini:2014zxa}, among others \cite{Ishii:2005xq, Gallouin:2012kb, Mundim:2013vca, Zlochower:2015baa}. Our intention is to provide a complementary analysis which extends recent GSF work on the dipolar (spin precession) and quadrupolar (tidal) sectors. We aim for completeness, by (i) seeking a complete basis of octupolar invariants, (ii) providing both numerical GSF data and high-order PN expansions at $\mathcal{O}(\mu/M)$.

In outline, the route to obtaining invariants is straightforward: (1) in the GSF formulation, the motion of a small compact body is associated with a geodesic in a \emph{regularly-perturbed vacuum} spacetime \cite{Detweiler:2002mi, Harte:2011ku}; (2) the electric tidal tensor $\EE_{ab}$ of the regularly-perturbed spacetime defines an orthonormal triad at each point on the geodesic; (3) the covariant derivative of the Riemann tensor $R_{abcd;e}$ resolved in this triad gives a set of well-defined scalar quantities $\{\chi_i\}$;  (4) the functional relationships $\chi_i(\Omega)$, where $\Omega$ is the circular-orbit frequency, are free of gauge ambiguities; (5) we define the `invariants' $\Delta \chi_i(\Omega)$ to be the $\mathcal{O}(\mu)$ parts of the differences $\chi_i(\Omega, \mu) - \chi_i(\Omega, \mu=0)$. 

The article is organized as follows. In Sec.~\ref{sec:formulation}, we introduce electric and magnetic tidal tensors of octupolar order; decompose in the `electric quadrupole' triad; examine the `background' ($\mu=0$) quantities; and apply perturbation theory to derive invariant quantities through $\mathcal{O}(\mu)$. In Sec.~\ref{sec:computational_approaches} we describe various computational approaches for obtaining the regular metric perturbation $h^R_{ab}$ and its associated invariants. In Sec.~\ref{sec:results} we present our results, primarily in the form of tables of data and PN series. 
In Sec.~\ref{sec:applications} we outline two wider applications of our work. We conclude with a discussion of progress and future work in Sec.~\ref{sec:conclusions}.

\emph{Conventions:}  We set $G = c = 1$ and use the metric signature $+2$. In certain contexts where the meaning is clear we also adopt the convention that $M = 1$. General coordinate indices are denoted with Roman letters $a,b,c, \ldots$, indices with respect to a triad are denoted with letters $i, j, k, \ldots$, and the index $0$ denotes projection onto the tangent vector. The coordinates $(t, r, \theta, \phi)$ denote general polar coordinates which, on the background Kerr spacetime, correspond to Boyer-Lindquist coordinates. Covariant derivatives are denoted using the semi-colon notation, e.g., $k_{a;b}$, with partial derivatives denoted with commas. Symmetrization and anti-symmetrization of indices is denoted with round and square brackets, $()$ and $[]$, respectively. 

% 

%Some references:
%\cite{Johnson-McDaniel:2015vva,
%Shah:2015nva,
%Kavanagh:2015lva,
%Shah:2015nva,
%Akcay:2015pza,
%Bini:2015mza,
%Dolan:2014pja,
%Isoyama:2014mja,
%Dolan:2013roa,
%Bini:2014zxa,
%Bini:2014ica,
%Poisson:2014gka,
%Tiec:2014lba,
%Merlin:2014qda,
%Shah:2014tka,
%Fujita:2012cm,
%Akcay:2012ea,
%LeTiec:2011bk,
%Barack:2011ed,
%Blanchet:2010zd,
%Barack:2010ny,
%Barack:2010tm,
%Barack:2009ey,
%Sago:2008id,
%Detweiler:2005kq,
%Detweiler:2000gt,
%Whiting:2003tx,
%Detweiler:2002mi}

\section{Formulation\label{sec:formulation}}

\subsection{Fundamentals}

\subsubsection{Tidal tensors}
We begin by considering a circular-orbit geodesic in the equatorial plane of the regularly-perturbed vacuum Kerr spacetime $g_{ab}$ with a tangent vector $u^a$. From the Riemann tensor $R_{abcd}$ (equal to the Weyl tensor $C_{abcd}$ in vacuum) we can construct electric-type and magnetic-type `quadrupolar' tensors,
\begin{eqnarray}
\EE_{ab} &=& R_{a c b d} u^c u^d, \\
\BB_{ab} &=& R^*_{a c b d} u^c u^d,  
\end{eqnarray}
where $R^*_{a  b c d}  = \frac{1}{2} \tensor{\varepsilon}{_{ab} ^{ef}} R_{efcd}$. We may also construct `octupolar' tensors,
\begin{eqnarray}
\EE_{a b c} &=& R_{a d b e ; c} u^d u^e , \\
\BB_{a b c} &=& R^*_{a d b e ; c} u^d u^e .
\end{eqnarray}
The quadrupolar tensors are symmetric ($\EE_{ab} = \EE_{ba}$, $\BB_{ab} = \BB_{ba}$), transverse ($\EE_{ab}u^b = 0 = \BB_{ab}u^b$), and traceless ($\tensor{\BB}{^a _a} = 0$ in general, $\tensor{\EE}{^a _a} = 0$ in vacuum). Similarly, the octupolar tensors are symmetric, and traceless in the first two indices (as $R_{a b c d} = R_{c d a b}$ and $R_{a b} = \tensor{R}{^c _{a c b}} = 0$) in vacuum. By contracting the Bianchi identity (or its dual)
$\tensor{R}{_{a b c d ; e}} + \tensor{R}{_{a b d e ; c}} + \tensor{R}{_{a b e c ; d}} = 0$, we observe that the octupolar tensors are also traceless on the latter pair of indices, $\tensor{\EE}{_{a b} ^{b}} = 0 = \tensor{\BB}{_{a b} ^{b}}$, in vacuum.  Note however that the octupolar tensors are {\it not} symmetric in the latter pair of indices, in general.

\subsubsection{Tetrad components\label{subsec:tetrad}}
Let us now introduce an orthonormal tetrad $\{e_0^a = u^a, e_i^\alp\}$ on the worldline and define tetrad-resolved quantities in the obvious way, so that
\beq
\XX_{i 0 j \ldots} = \XX_{a b c \ldots} e_i^a u^b e_j^c \ldots  , 
\eeq
where $\XX_{abc \ldots}$ is any tensor and $i,j,k \in \{1,2,3\}$. The quadrupole components are spatial, $\EE_{00} = \EE_{0i} = 0 = \BB_{0i} = \BB_{00}$. The octupole components are spatial in first two indices, but not in general. We may then consider three types of octupolar terms, namely,
\beq
\EE_{i j 0} ,  \quad \EE_{i [j ; k]}, \quad \text{and} \quad \EE_{(i j k)} , \label{eq:three-types}
\eeq
and similarly for $\BB$. Here $()$ and $[]$ denotes the complete symmetrization and anti-symmetrization of indices.   

Note that $\EE_{ij}$ is real and symmetric, and thus its eigenvalues are real and its eigenvectors are orthogonal.  Thus, we may select our triad $e_a^i$ to coincide with the electric-quadrupolar eigenbasis. In other words, we choose the triad in which $\EE_{ij}$ is diagonal. We choose $e_2^a$ to be the vector orthogonal to the equatorial plane.

\subsubsection{Equatorial symmetry}
For circular equatorial orbits, the reflection-in-equatorial plane symmetry implies that many components are identically zero. Namely,
\begin{eqnarray} &
\EE_{12} = \EE_{23} = \BB_{11} = \BB_{13} = \BB_{22} = \BB_{33} = 0 , \\
&
\EE_{112} = \EE_{222} = \EE_{233} = \EE_{123} = 0 , \\
&
\BB_{111} = \BB_{122} = \BB_{133} = \BB_{113} = \BB_{223} = \BB_{333} = 0 ,
\end{eqnarray}
with all permutations of these indices also zero. 

\subsection{Classification of octupolar components\label{subsec:components}}
Now we consider the three types of terms (\ref{eq:three-types}) separately, and show that $\XX_{ij0}$ and $\XX_{i[jk]}$ may be derived from dipolar and quadrupolar terms, whereas $\XX_{(ijk)}$ encode new information at octupolar order. 

\subsubsection{$\EE_{i j 0}$ and $\BB_{i j 0}$}
For circular orbits, we have $u^b e^a_{1;b} = \omega \, e_3^a$, $u^b e^a_{2;b} = 0$ and $u^b e^a_{3;b} = - \omega \, e_1^a$ where $\omega$ is the precession frequency with respect to proper time, defined by parallel transport observed from the electric eigenbasis (c.f.~Ref.~\cite{Dolan:2013roa, Dolan:2014pja}). As the quadrupolar eigenvalues are time-independent on circular orbits, the only non-trivial components are
\beq
\EE_{1 3 0} = \omega \left( \EE_{11} - \EE_{33} \right), \quad
\BB_{1 2 0} = - \omega \, \BB_{23} , \quad \BB_{2 3 0} = \omega \, \BB_{12} .
\label{eq:Eij0}
\eeq
% Note that (i) $\BB_{23} = 0$ on the background spacetime, and (ii) $\Delta \BB_{23} = - \bar{\lambda}^B \Delta \chi$, where $\Delta \chi$ was computed in our quadrupolar paper (this follows from text above Eq.~(2.54) in the quadrupolar paper, as we are now choosing the tetrad to be aligned with the electric eigenbasis). 
%

\subsubsection{$\EE_{i [j k]}$ and $\BB_{i [j k]}$}
By virtue of the the Bianchi identity,
\beq
\EE_{a [b c]} = -\frac{1}{2} u^e \left( u^d R_{d a b c} \right)_{; e}  , \quad \quad
\BB_{a [b c]} = -\frac{1}{2} u^e \left( u^d R^\ast_{d a b c} \right)_{; e}  .
\eeq
We now (i) project onto the tetrad, (ii) use that $\BB_{i j} = \frac{1}{2} \eps_{j k l} R_{0 i k l}$ and $\EE_{i j} = - \frac{1}{2} \eps_{j k l} R^\ast_{0 i k l}$, and (iii) recall that the tetrad components in the electric frame are constants for circular orbits. Thus all components are zero except
\begin{eqnarray}
\EE_{2 [2 3]} =  \EE_{1 [3 1]}  &=& \phantom{-} \frac{1}{2} \omega \BB_{23} , \label{eq:E131} \\
\EE_{3 [3 1]}  = \EE_{2 [1 2]}  &=& - \frac{1}{2} \omega \BB_{12} , \\
\BB_{1 [1 2]} = \, \BB_{3 [2 3]} &=& \phantom{-} \frac{1}{2} \omega \left( \EE_{11} - \EE_{33} \right) ,  \label{eq:B323} 
\end{eqnarray}
and permutations thereof.

\subsubsection{$\EE_{(i j k)}$ and $\BB_{(i j k)}$}
In general, $\EE_{(i j k)}$ and $\BB_{(i j k)}$ each have ten components satisfying 3 trace conditions, i.e., seven independent components each. For circular orbits, 4 electric and 6 magnetic components are zero, respectively, leaving 6 and 4 non-trivial quantities satisfying 2 and 1 non-trivial gauge constraints. In other words, there are 10 quantities we may calculate (given below), satisfying 3 non-trivial trace conditions; thus, 7 new independent degrees of freedom at octupolar order.

\subsubsection{Additional invariants}
Other octupolar quantities may be written in terms of the set identified above. For example, a relevant quantity in EOB theory [see Ref.~\cite{Bini:2014zxa}, Eq.~(D10)] is $K_{3+} \equiv  \EE_{(abc)} \EE^{(abc)}$, which may be expressed as
\beq
K_{3+} = \EE_{(111)}^2 + \EE_{(333)}^2 + 3 \left( \EE_{(122)}^2 + \EE_{(133)}^2 + \EE_{(311)}^2 + \EE_{(322)}^2 \right) - 6 \EE_{(130)}^2 ,  \label{eq:K3p-def}
\eeq
where $\EE_{(130)} = \frac{1}{3} \EE_{130}$.

\subsection{Circular orbits: Background quantities\label{subsec:testparticle}}
Below we give the values of the tidal quantities for circular equatorial geodesics on the unperturbed Kerr spacetime, i.e., for test-masses ($\mu=0$). Here, the orbital radius is $r_0$ and the orbital frequency is $\Omega = \sqrt{M} / (r_0^{3/2} + a \sqrt{M})$ where $a$ is the Kerr spin parameter and $a > 0$ ($a < 0$) for prograde (retrograde) orbits.  % Here we set $M=1$ for convenience.

The tangent vector $u^a$ and electric-eigenbasis triad have the components
\cite{Marck:1983}
\begin{subequations}
\begin{eqnarray}
u^a &=& [U , 0, 0, \Omega U], \\
e_1^a &=& [0, \sqrt{\Delta_0} / r_0, 0, 0] , \label{rhodef} \\
e_2^a &=& [0,0,1/r_0,0], \\
e_3^a &=& -\tensor{\epsilon}{^a _{b c d}} u^b e_1^c e_2^d,
\end{eqnarray}
\end{subequations}
where $U = \sqrt{M} / (\Omega r_0^{3/2} \upsilon)$, $\Delta_0 = r_0^2 - 2Mr_0 + a^2$ and
\beq
\upsilon^2 \equiv 1 - 3M/r_0 + 2 a \sqrt{M} / r_0^{3/2}.   \label{eq:upsilon}
\eeq
The spin precession rate is $\omega = \sqrt{M r_0} / r_0^{2}$.

\subsubsection{Quadrupolar components}
 The (non-trivial) quadrupolar components are 
\begin{subequations}
\begin{eqnarray}
\EE_{11} &=& \frac{M}{r_0^3} - \frac{3 M \Delta_0}{\upsilon^2 r_0^5}, \\
\EE_{22} &=& -\frac{2M}{r_0^3} + \frac{3 M \Delta_0}{\upsilon^2 r_0^5}, \\
\EE_{33} &=& \frac{M}{r_0^3} , \\
\BB_{12} &=& -\frac{3 M^{3/2} \sqrt{\Delta_0}  \left(1 - a/\sqrt{M r_0} \right)}{r_0^{9/2} \upsilon^2} ,
\end{eqnarray}
\end{subequations}
We note that  $\BB_{23} = 0$ on the background.

\subsubsection{Octupolar components}
In the electric sector,
\begin{eqnarray}
\EE_{(111)} &=& + \mathcal{A} \left( 6 r_0^2 - 9 M r_0 - 12 a \sqrt{Mr_0} + 15 a^2 \right) \label{eq:E111:background} \\
\EE_{(122)} &=& - \mathcal{A} \left( 3r_0^2 - 2M r_0 - 16 a \sqrt{Mr_0} + 15 a^2 \right) \\
\EE_{(133)} &=& - \mathcal{A} \left( 3 r_0^2 - 7Mr_0  + 4 a \sqrt{Mr_0} \right), 
\end{eqnarray}
where $\mathcal{A} = \sqrt{\Delta_0} M / (r_0^7 \upsilon^2)$. We note that $\EE_{(311)} = \EE_{(322)} = \EE_{(333)} = 0$ on the background.

In the magnetic sector, 
\begin{eqnarray}
\BB_{(211)} &= + \mathcal{C} \left(4 r_0^2 - 8 M r_0 + 7 a^2\right) & - \mathcal{D} \left( 4r_0^2 - 7Mr_0 + 5a^2\right) , \\
\BB_{(222)} &= -\mathcal{C} \left(3 r_0^2 - 6 M r_0 + 9 a^2\right) & + \mathcal{D} \left( 3r_0^2 - 4Mr_0 + 5a^2\right) , \\
\BB_{(233)} &= -\mathcal{C} \left(\phantom{x} r_0^2 - 2M r_0 - 2a^2\right) & + \mathcal{D} \left(\phantom{x} r_0^2 - 3M r_0 \right) , \label{eq:B233:background}
\end{eqnarray}
where $\mathcal{C} = 2M^{3/2} / (r_0^{13/2} \upsilon^2)$ and $\mathcal{D} = 3 aM / (r_0^7 \upsilon^2)$. Note that $\BB_{(123)} = 0$ on the background. 

The `derived' quantities $\EE_{i j 0}$, $\EE_{i [j k]}$, etc., may be easily calculated using Eq.~(\ref{eq:Eij0}), Eq.~(\ref{eq:E131})--(\ref{eq:B323}) and Eq.~(\ref{eq:E111:background})--(\ref{eq:B233:background}). For example, in the Schwarzschild ($a=0$) case, using Eq.~(\ref{eq:K3p-def}) yields
%Using $\EE_{(130)} = - \frac{3f}{r^{9/2} \upsilon^2}$
%the Schwarzschild value of $K_{3+}$ is
\beq
K_{3+} = \frac{6M^2(1-2M/r_0) (15 r_0^2 - 46 M r_0 + 42 M^2)}{r_0^{10} (1 - 3M/r_0)^2} .  \label{eq:K3p-schw}
\eeq

\subsection{Circular orbits: Perturbation theory}
Here we seek expressions for the octupolar quantities in the regular perturbed spacetime $\bar{g}_{ab} + h_{ab}^R$, where $\bar{g}_{ab}$ is the Kerr metric in Boyer-Lindquist coordinates, and $h_{ab}^R = \mathcal{O}(\mu)$ is the `regular' metric perturbation defined by Detweiler \& Whiting \cite{Detweiler:2002mi}. We work to first order in the small mass $\mu$, neglecting all terms at $\mathcal{O}(\mu^2)$, and noting that the regular perturbed spacetime is Ricci-flat.

We take the standard two-step approach \cite{Detweiler:2008ft, Sago:2008id, Dolan:2014pja}. For a given geodesic quantity $\chi$ (e.g.~$\EE_{(111)}$), we first compare $\chi$ on a circular geodesic in the perturbed spacetime with $\chi$ on a circular geodesic the background spacetime at the same coordinate radius $r=r_0$. %Here, we are implicitly using a diffeomorphism that associates points with the same coordinates.
Then, noting that $r_0$ itself varies under a gauge transformation at $\mathcal{O}(\mu)$, we apply a correction to compare $\chi$ on two geodesics which share the same orbital frequency $\Omega$. 

Following the convention of Ref.~\cite{Dolan:2014pja}, we use an `over-bar' to denote `background' quantities, so that barred quantities such as $\bar{u}^a$ are assigned the same coordinate values as in Sec.~\ref{subsec:testparticle}. We use $\delta$ to denote the difference at $\mathcal{O}(\mu)$, i.e.,~$\delta e^a_i \equiv e^a_i - \bar{e}^a_i$. At $\mathcal{O}(\mu)$, $\delta$ may be applied as an operator with a Leibniz rule $\delta (AB) = (\delta A)B + A \delta B$. In general, such differences are gauge-dependent. To obtain an invariant difference, we introduce the `frequency-radius' $r_{\Omega}$ via
\beq
\Omega = \sqrt{M} / (r_\Omega^{3/2} + a \sqrt{M}) . \label{eq:freqradius}
\eeq
Then, we write
\beq
\chi(r_{\Omega}) - \bar{\chi}(r_{\Omega}) = \Delta \chi(r_0) + \mathcal{O}(\mu^2).  \label{eq:Deltadef}
\eeq
Here $\bar{\chi}(r_\Omega)$ has the same functional form as $\chi$ on the background spacetime, with $r_0$ replaced by $r_{\Omega}$. As $\Delta \chi$ is at $\mathcal{O}(\mu)$, we may parameterize $\Delta \chi$ using the $\mathcal{O}(\mu^0)$ `background' radius $r_0$, rather than $r_{\Omega}$, as $r_0 - r_{\Omega} = \mathcal{O}(\mu)$. Such relationships, $\Delta \chi(r_0)$, are invariant within the class of gauges in which the metric perturbation is helically-symmetric (implying that $\bar{u}^c h^R_{a b , c} = 0$ at the relevant order).

\subsubsection{Perturbation of the tetrad}
We may write the variation of the tetrad legs in the following way,
\begin{subequations}
\begin{eqnarray}
\delta u^a &=& \beta_{00} \bar{u}^a + \beta_{03} \bar{e}_3^a , \\
\delta e^a_i &=&  \beta_{i 0} \bar{u}^a + \sum_{j=1}^3 \beta_{ij} \bar{e}_j^a .
\end{eqnarray}
\end{subequations}
with the coefficients $\beta_{ab} = \mathcal{O}(\mu)$ to be determined below. First, we note that $\beta_{00}$ and $\beta_{03}$ may be found by recalling key relations previously established in GSF theory for equatorial circular orbits on Kerr
spacetime \cite{Detweiler:2008ft, Shah:2012gu}, namely,
\begin{eqnarray}
\frac{\delta u^t}{\bar{u}^t} &=& \frac{1}{2} h_{00} - \frac{\bar{\Omega}}{2} \sqrt{\frac{r_0}{M}} \left( r_0^2 + a^2 - 2 a\sqrt{Mr_0} \right) \tilde{F}_r ,   \label{eq:ut} \\
\frac{\delta u^\phi}{\bar{u}^\phi} &=&  \frac{1}{2} h_{00} - \frac{1}{2M} \left( r_0^2 - 2M r_0 + a \sqrt{M r_0} \right) \tilde{F}_r  .\label{eq:uphi}
\end{eqnarray}
Here $h_{00} \equiv h^R_{a b} \bar{u}^a \bar{u}^b$, and the radial component of the GSF is given by
\beq
\tilde{F}_r \equiv \mu^{-1} F_r =  \left. \frac{1}{2} \bar{u}^a \bar{u}^b \frac{\partial h^R_{a b}}{\partial r}  \right|_{r=r_0}.
\eeq
Hence we have $\beta_{00} = \frac{1}{2} h_{00}$ and $\beta_{03} = -\frac{1}{2} \sqrt{\frac{r_0 \Delta_0}{M}} \, \tilde{F}_r$, where $h_{00} = h^{R}_{ab} \bar{u}^a \bar{u}^b$ and $\tilde{F}_r = \mu^{-1}F_r = \mathcal{O}(\mu)$ is the (specific) radial self-force. The diagonal coefficients $\beta_{ii}$ follow from the normalization condition, $\left(\bar{g}_{ab} + h^R_{ab}\right) \left(\bar{e}_i^a + \delta e_i^a\right) \left(\bar{e}_j^b + \delta e_j^b\right) = \delta_{ij}$. That is, $\beta_{ii} = -\frac{1}{2} h_{ii}$, where $h_{ii} = h^R_{ab} \bar{e}^a_i \bar{e}^b_i$ (no summation implied). From orthogonality of legs $0$ and $3$, we have $\beta_{30} = \beta_{03} + h_{03}$ where $h_{03} = h^R_{ab} \bar{u}^a \bar{e}^b_3$. By similar reasoning, $\beta_{10} = h_{01}$ and $\beta_{31} + \beta_{13} + h_{13} = 0$. To eliminate the residual rotational freedom in the triad at $\mathcal{O}(\mu)$, we now impose the condition that the triad is aligned with the electric eigenbasis, i.e., that $\EE_{ij}$ is diagonal in the perturbed spacetime (so that, e.g., $\EE_{13} = 0$). From this condition it follows that
\begin{eqnarray}
\beta_{13} &=& \frac{(\delta R)_{1030} - \bar{\EE}_{11} h_{13} }{ \bar{\EE}_{11} - \bar{\EE}_{33} } , \\
\beta_{31} &=& \frac{-(\delta R)_{1030} + \bar{\EE}_{33} h_{13} }{ \bar{\EE}_{11} - \bar{\EE}_{33} } ,
\end{eqnarray}
where 
\begin{equation}
\left(\delta R\right)_{1030} = \delta R_{abcd} \, \bar{e}_1^a \bar{u}^b \bar{e}_3^c \bar{u}^d .
\end{equation}

\subsubsection{Perturbation of octupolar components}
\label{sec:pert-octupole}
Here we present results for the perturbation of the (symmetrized) octupolar components $\EE_{(ijk)}$ and $\BB_{(ijk)}$. The electric components are
\begin{subequations}
\begin{eqnarray}
\delta \EE_{(111)} &=& (\delta \nabla R)_{(10101)} + \left(h_{00} - \frac{3}{2}h_{11}\right) \bar{\EE}_{(111)} + 2 \beta_{03} \RR_{10131} ,  \label{Eijk:first} \\
\delta \EE_{(122)} &=& (\delta \nabla R)_{(10202)} + \left(h_{00} - \frac{1}{2}h_{11} - h_{22} \right) \bar{\EE}_{(122)} + 2 \beta_{03} \RR_{10232} , \\
\delta \EE_{(133)} &=& (\delta \nabla R)_{(10303)} + \left(h_{00} - \frac{1}{2}h_{11} - h_{33} \right) \bar{\EE}_{(133)} + 2 \beta_{03} \RR_{10333} + \frac{2}{3} \beta_{30} \, \bar{\omega} \left(\bar{\EE}_{11} - \bar{\EE}_{33}\right) , \\
\delta \EE_{(113)} &=& (\delta \nabla R)_{(10103)} + \frac{2}{3} \beta_{10} \bar{\omega} \left(\bar{\EE}_{11} - \bar{\EE}_{33}\right) + \beta_{31} \bar{\EE}_{(111)} + 2 \beta_{13} \bar{\EE}_{(133)} , \\
\delta \EE_{(223)} &=& (\delta \nabla R)_{(20203)} + \beta_{31} \bar{\EE}_{(122)} , \\
\delta \EE_{(333)} &=& (\delta \nabla R)_{(30303)} + 3 \beta_{31} \bar{\EE}_{(133)} .  \label{Eijk:last}
\end{eqnarray}
\label{eq:Eijk}
\end{subequations}
where
\begin{eqnarray}
(\delta \nabla R)_{(i0j0k)} &=& \delta R_{a b c d ; e} \bar{u}^{b} \bar{u}^{d} \bar{e}_i^{(a} \bar{e}_j^{c} \bar{e}_k^{e)}, \\
\RR_{i0j3k} &=& \bar{R}_{a b c d ; e} \bar{u}^{(b} \bar{e}_3^{d)} \bar{e}_i^{(a} \bar{e}_j^{c} \bar{e}_k^{e)} .
\end{eqnarray}
The magnetic components are
\begin{subequations}
\begin{eqnarray}
(\delta \BB)_{(211)} &=& (\delta \nabla R^\ast)_{20101} + \left(h_{00} - h_{11} - \frac{1}{2} h_{22} \right) \bar{\BB}_{(211)} + 2 \beta_{03} \RR^\ast_{20131} ,  \label{Bijk:first}  \\
(\delta \BB)_{(222)} &=& (\delta \nabla R^\ast)_{20202} + \left(h_{00} - \frac{3}{2}h_{22}\right) \bar{\BB}_{(222)} + 2\beta_{03} \RR^\ast_{20232} , \\
(\delta \BB)_{(233)} &=& (\delta \nabla R^\ast)_{20303} + \left(h_{00} - \frac{1}{2}h_{22} - h_{33} \right) \bar{\BB}_{(233)} + 2\beta_{03} \RR^\ast_{20333} + \frac{2}{3} \beta_{30} \bar{\omega} \bar{\BB}_{12} , \\
(\delta \BB)_{(123)} &=& (\delta \nabla R^\ast)_{10203} + \beta_{13} \bar{\BB}_{(233)} + \beta_{31} \bar{\BB}_{(211)} + \frac{1}{3} \beta_{10} \bar{\omega} \bar{\BB}_{12} . \label{Bijk:last} 
\end{eqnarray}
\label{eq:Bijk}
\end{subequations}
where
\begin{eqnarray}
(\delta \nabla R^\ast)_{(i0j0k)} &=& \delta R^\ast_{a b c d ; e} \bar{u}^{b} \bar{u}^{d} \bar{e}_i^{(a} \bar{e}_j^{c} \bar{e}_k^{e)} , \\
\RR^\ast_{i0j3k} &=& \bar{R}^\ast_{a b c d ; e} \bar{u}^{(b} \bar{e}_3^{d)} \bar{e}_i^{(a} \bar{e}_j^c \bar{e}_k^{e)} .
\end{eqnarray}

\subsubsection{Invariant relations}
As noted above, the coordinate radius of the orbit, $r=r_0$, is not invariant under changes of gauge (i.e., coordinate changes at $\mathcal{O}(\mu)$). On the other hand, the orbital frequency $\Omega$ is invariant under helically-symmetric gauge transformations. Following Eq.~(\ref{eq:Deltadef}), we may therefore express the functional relationship between $\chi \in \left\{ \EE_{(111)}, \ldots \right\}$ and $\Omega$ as follows,
\beq
\chi(r_\Omega) = \bar{\chi}(r_{\Omega}) + \Delta \chi(r_0) + \mathcal{O}(\mu^2),
\eeq
where $r_{\Omega}$ is the frequency-radius defined in Eq.~(\ref{eq:freqradius}), and $\Delta \chi = \mathcal{O}(\mu)$. Note that $\bar{\chi}(r_\Omega)$ denotes the `test-particle' functions defined in Sec.~\ref{subsec:testparticle} evaluated at $r_\Omega$. By definition, we have $\Delta \Omega = 0$. At $\mathcal{O}(\mu)$,
\beq
\Delta \chi = \delta \chi - \delta \Omega \frac{d r_0}{d \bar{\Omega}} \frac{d \bar{\chi}}{d r_0} ,
\eeq
or, making use of Eq.~(\ref{eq:ut}) and (\ref{eq:uphi}) and $\delta \Omega / \bar{\Omega} = \delta u^\phi / \bar{u}^\phi - \delta u^t / \bar{u}^t$,
\beq
\Delta \chi = \delta \chi - \frac{1}{3M} r_0^3 \upsilon^2 \tilde{F}_r \frac{d\bar{\chi}}{dr_0} . \label{dY}
\eeq
In summary, $\Delta \chi$ defined by Eq.~(\ref{dY}), Eq.~(\ref{eq:Eijk}) and Eq.~(\ref{eq:Bijk}) are the invariant quantities which we will compute in the next sections.

\subsubsection{Further quantities}
In Sec.~\ref{subsec:components} we wrote $\EE_{i j 0}$, $\EE_{i [j k]}$, $\BB_{ij0}$, $\BB_{i [j k]}$ in terms of quadrupolar tidal components, and the spin precession scalar $\omega$. If required, one may deduce the variation of these components by applying $\Delta$ as a Leibniz operator.  For example, starting with Eq.~(\ref{eq:Eij0}), 
\beq
\Delta \EE_{130} = \Delta \omega \left( \bar{\EE}_{11} - \bar{\EE}_{33} \right) + \bar{\omega} \left( \Delta \EE_{11} - \Delta \EE_{33} \right) .
\eeq
Numerical data for the variation in the quadrupolar components $\Delta \EE_{11}, \ldots, \Delta \BB_{21}, \ldots$ is given in Table I of Ref.~\cite{Dolan:2014pja}. We may compute $\Delta \omega$ from the redshift and spin-precession invariants, $\Delta U$ and $\Delta \psi$, using
\begin{equation}
\Delta \omega = \frac{\bar{\omega}}{\bar{U}}\Delta U -  \bar{U}\bar{\Omega} \Delta \psi ,
\end{equation}
together with the data in Table III of Ref.~\cite{Dolan:2014pja}. 

Similarly, the variation $\Delta K_{3+}$, for example, can be found by applying
$\Delta$ in this manner to Eq.~(\ref{eq:K3p-def}). This can then be related to
the quantity $\hat{\delta}_{K3+}$, whose post-Newtonian expansion was given
to $7.5\text{PN}$ in Ref.~\cite{Bini:2014zxa}. %(we correct an error in the order $y^6$ term and extend the expansion to higher order in this work). 
Noting that
$K_{3+} \equiv \Gamma^4 \mathcal{K}_{3+}$ and
\begin{equation}
  \Gamma = \frac{1}{\sqrt{1 - 3 M / r_0}} \bigg[1 + \frac{1}{2} h_{00} + \mathcal{O}(\mu^2) \bigg],
\end{equation}
we then have a relation between the first-order perturbations,
\begin{equation}\label{eq:hatdeltaK3p}  
\frac{\Delta K_{3+}}{\bar{K}_{3+}} = \hat{\delta}_{K3+} + 2 h_{00}.
\end{equation}

\section{Computational approaches}\label{sec:computational_approaches}

In this section we outline our methods for computing the octupolar invariants for a particle of mass $\mu$ on a circular orbit of radius $r_0$ in Schwarzschild geometry. Our approaches break into two broad catagories: (i) numerical integration of the linearized Einstein equation in either the Regge-Wheeler (RW) or Lorenz gauge and (ii) analytically solving the Regge-Wheeler field equations as a series of special functions via the Mano-Suzuki-Takasugi (MST) method. In both cases we decompose the linearized Einstein equation into tensor-harmonic and Fourier modes and solve for the resulting decoupled radial equation. In this section, and subsections that follow, $l$ and $m$ are the tensor-harmonic multipole indices, $\omega$ is the mode frequency and we work with standard Schwarzschild coordinates $(t,r,\theta,\varphi)$. For this section let us also define $f\equiv f(r) = 1-2M/r$. We shall also use a subscript `0' to denote a quantity evalulated at the particle. Finally, note that for a circular orbit about a Schwarzschild black hole the particle's (specific) orbital energy and angular-momentum are given by 
\begin{align}
	\en	=	\frac{r_0-2M}{\sqrt{r_0(r_0-3M)}},		\qquad 		\ang = r_0\sqrt{\frac{M}{r_0-3M}},
\end{align}
respectively.

For calculations in the RW gauge there is a single `master' radial function, $\Psi_{lm\omega}$, to be solved for each tensor-harmonic and Fourier mode \cite{Regge-Wheeler,Zerilli}. For circular orbits the Fourier spectrum is discrete and given by $\omega\equiv\omega_m = m\Omega$ where $\Omega=\sqrt{M/r_0^3}$ is the azimuthal orbital frequency. Consequently, we label the RW master function with only $lm$ subscripts hereafter. The full metric perturbation can be rebuilt from the $\Psi_{lm}$'s and their derivatives \cite{Berndtson:thesis}. For $l\ge2$ the ordinary differential equation that $\Psi_{lm}$ obeys takes the form
\begin{align}\label{eq:RW}
	\left(\sdiff{}{r_*}  + \left[\omega_m^2 - U_l(r) \right]\right) \Psi_{lm} = \mathcal{S}_1\delta(r-r_0) + \mathcal{S}_2\delta'(r-r_0),
\end{align}
where $r_*$ is the radial `tortoise' coordinate given by $dr_*/dr=f^{-1}$ and $U(r)$ is an effective potential. The effective potential used depends on whether the perturbation is odd or even parity. For the odd/even parity modes, equivalently $l+m=\text{odd/even}$, the potential is given by
\begin{align}
	U^o_l(r) &= \frac{f}{r^2}\left(l(l+1) -\frac{6M}{r}\right),		\\
	U^e_l(r) &= \frac{f}{r^2\Lambda^2}\left[ 2\lambda^2\left( \lambda+1+\frac{3M}{r}\right) + \frac{18M^2}{r^2}\left(\lambda+\frac{M}{r}\right) \right],
\end{align}
respectively, where $\lambda=(l+2)(l-1)/2$ and $\Lambda = \lambda+3M/r_0$. The form of the source terms, $\mathcal{S}_i$, also differ for the even and odd sectors. Explicitly, the odd sector sources take the form \cite{Hopper:2010uv}
\begin{align}\label{eq:Odd_Source}
\mathcal{S}_1^\text{o}&=-\frac{2 p  f_0\mathcal{L}_0}{\lambda l(l+1)}X_\phi^*(\theta,\phi),\\
\mathcal{S}_2^\text{o}&=\frac{2 p r_0 f_0^2  \mathcal{L}_0}{\lambda l(l+1)}X_\phi^*(\theta,\phi).
\end{align}
For the even sector, we have
\begin{align}\label{eq:Even_Source}
	\mathcal{S}_1^\text{e} &=\frac{p q \mathcal{E}_0}{r_0 f_0 \Lambda}\left[\frac{\mathcal{L}_0^2}{\mathcal{E}_0^2} f_0^2 \Lambda -(\lambda(\lambda+1)r_0^2+6\lambda M r_0+15 M^2) \right]Y_{lm}^*(\theta,\phi) -\frac{4 p \mathcal{L}_0^2 f_0^2}{r_0\mathcal{E}_0}\frac{(l-2)!}{(l+2)!}Y_{\phi \phi}^*(\theta,\phi), \\
	\mathcal{S}_2^\text{e}&=( r_0^2 p q \mathcal{E}_0) Y^*_{lm}(\theta,\phi),
\end{align}
where we have defined the following expressions for convenience:
\begin{align}
	X_\phi(\theta,\phi)			&= \text{sin} \theta \partial_\theta Y_{lm} (\theta,\phi),			\\
	Y_{\phi \phi}(\theta,\phi)	&=\big(\partial_{\phi\phi} +\text{sin}\theta \text{cos}\theta \partial_\theta +\frac{l(l+1)}{2}\text{sin}^2\theta\big)Y_{lm}(\theta,\phi)\\
	p							&=\frac{8 \pi \mu}{r_0^2}, \qquad q=\frac{f_0^2}{(\lambda+1)\Lambda}.
\end{align}

For the radiative modes $l\ge2, m\neq0$ we will construct homogeneous solutions to Eq.~\eqref{eq:RW} either numerically or as a series of special functions, as outlined in the subsections below. For the static ($l\ge2$,$m=0$) modes, closed-form analytic solutions to the homogeneous RW equation are known. In the odd sector these can be written in terms of standard hypergeometric functions:
\begin{align}
	\tilde{\Psi}^{o-}_{l0} 		&= x^{-l-1} {}_2F_1(-l-2,-l+2,-2l,x)		\label{eq:RW_static_odd_in}					\\
	\tilde{\Psi}^{o+}_{l0}		&= x^l {}_2F_1(l-1,l+3,2+2l,x).										\label{eq:RW_static_odd_out}			
\end{align}
where hereafter an overtilde denotes a homogeneous solution, a `$+$' superscript denotes an outer solution (regular at spatial infinity, divergent at the horizon), a `$-$' denotes an inner solution (regular at the horizon, divergent at spatial infinity) and $x=2M/r$. In practice, we need only solve the simpler odd sector field equations, and construct the even sector homogeneous solutions via the transformation \cite{Berndtson:thesis}:
\begin{align}
\tilde{\Psi}^{e\pm}_{lm}	=& \frac{1}{\lambda+\lambda^2 \pm 3 i \omega M}\left[\left(\lambda+\lambda^2+\frac{9M^2(r-2M)}{r^2(r\lambda+3M)}\right)\tilde{\Psi}^{o\pm}_{lm} + 3Mf \frac{d\tilde{\Psi}^{o\pm}_{lm}}{dr}\right].	\label{eq:RW_to_Z}
\end{align}
Note this equation holds for both static and radiative modes.

We construct the inhomogeneous solutions to Eq.~\eqref{eq:RW} via the standard Variation of Parameters method. As the source contains both a delta-function and the derivative of a delta-function the inhomogeneous solution and its radial derivative will both be discontinuous at the particle. Constructing the inhomogeneous solutions then becomes a `matching' proceedure with the jump in the field and its derivative across the particle governed by coefficients $\mathcal{S}_1$ and $\mathcal{S}_2$. Supressing even/odd notation, we define matching coefficients as follows: 
\begin{align}
	D^{\pm}_{lm}=\frac{1}{W_{lm}}\left[\left(\frac{\mathcal{S}_1}{f_0}+\frac{2M\mathcal{S}_2}{r_0^2 f_0^2}\right)\Psi^{\mp}_{lm}-\frac{\mathcal{S}_2}{f_0}\partial_r \Psi^{\mp}_{lm}\right],
\end{align}
with the usual Wronskian defined as $W_{lm}=f_0(\tilde{\Psi}_{lm}^- \partial_r \tilde{\Psi}_{lm}^+ - \tilde{\Psi}_{lm}^+ \partial_r \tilde{\Psi}_{lm}^-)$. Finally we construct the inhomogeneous solutions via 
\begin{align}\label{eq:inhom_sols}
	\Psi_{lm}^{\pm}(r)=D_{lm}^{\pm} \tilde{\Psi}_{lm}^{\pm}(r).
\end{align}
where the $D_{lm}^{\pm}$'s are constants for all values of $r$.

To complete the metric perturbation in the RW gauge we use the $l=0$ and $l=1$ results of Zerilli \cite{Zerilli:1970}. Detweiler and Poisson expressed these contributions succinctly for circular orbits \cite{Detweiler-Poisson:2003}. For the monopole and static dipole we have
\begin{align}
	h_{tt}^{l=0} 			&= 2\mu\en\left(\frac{1}{r} + \frac{f}{r_0-2M}\right) \Theta(r-r_0),				\label{eq:RW_tt_monopole}						\\
	h_{rr}^{l=0} 			&= \frac{2\mu\en}{rf^2}\Theta(r-r_0),												\label{eq:RW_rr_monopole}												\\
	h_{t\varphi}^{l=1,m=0} 	&= -2\mu\ang\sin^2\theta \left\{\begin{array}{c c} r^2/r_0^3 & r<r_0 \\ 1/r & r>r_0 \end{array} \right.,	\label{eq:odd_dipole}		
\end{align}
where $\Theta$ is the Heaviside step function and all other components are zero. The $l=1,m=1$ mode does not contribute to our gauge invariant quantities so we will not give the explicit expression for the non-zero $h_{tt}$, $h_{tr}$ and $h_{rr}$ components of this mode (but as a check we use the expressions, given as Eqs.~(5.1)-(5.3) in Ref.~\cite{Detweiler-Poisson:2003}, to check that the contribution from this mode to our invariants is identically zero).

As well as working in the RW gauge we also make a computation in the Lorenz gauge. Our code is a \textit{Mathematica} re-implentation of that presented by Akcay \cite{Akcay:2010} and as such we refer the reader to that work for further details.

\subsection{Numerical computation of the retarded metric perturbation}

For our RW gauge calculuation, as discussed above, analytic solutions are known for the monopole, dipole and static ($m=0$) modes. This only leaves the radiative modes ($l\ge2, m \neq 0$) to be solved for numerically. Our numerical routines are implemented in \textit{Mathematica} which allows us to go beyond machine precision in our calculation with ease. Given suitable boundary conditions near the black hole horizon and at a sufficiently large radius (we discuss below how we choose these radii in practice),  we use Mathematica's \texttt{NDSolve} routine to solve for the inner and outer solutions to the homogeneous Regge-Wheeler equation \eqref{eq:RW}. Inhomogenous solutions are then constructed by imposing matching conditions of these functions at the location of the orbiting particle. 

\subsubsection{Numerical boundary conditions}

In order to construct boundary conditions, we use an appropriate power law ansatz for $\Psi_{RW}$ at each of our boundaries, given by 
\begin{align}
	\tilde{\Psi}_{RW}^\infty &\sim e^{i \omega r_*}\sum_{n=0}^{n_+}\frac{a_n}{(\omega r_\infty)^n}	,			\label{eq:horizon_expansion}	\\
	\tilde{\Psi}_{RW}^\text{H} &\sim e^{-i \omega r_*}\sum_{n=0}^{n_-}b_n f(r_H)^n	.						\label{eq:infinity_expansion}
\end{align}
Recursion relations for the series coefficients can be found by inserting our ansatz into the homogeneous RW equations, and choosing a maximum number of outer and inner terms $n_{\text{max}}=n_{\pm}$ gives us initial values for our fields at these boundaries. Inserting \eqref{eq:horizon_expansion} and \eqref{eq:infinity_expansion} into \eqref{eq:RW} for the odd sector, we find the following recursion relations:  
\begin{align}
	a_n &= \frac{i}{2n}\left[(l(l+1)-n(n-1))a_{n-1} + 2M\omega(n-3)(n-1)a_{n-2}\right],				\\
	b_n &= \frac{1}{n(n-4iM\omega)}\left[(l(l+1)+2n(n-1)-3)a_{n-1} - (n+1)(n-3)a_{n-2}\right].
\end{align}
As discussed above we do not need to solve the even sector field equations as we can transform from the simpler odd sector solutions using Eq.~\eqref{eq:RW_to_Z}. 

For the inner homogeneous solutions, the convergence of the series \eqref{eq:horizon_expansion} improves with increasing $n_{-}$, and in practice we choose $n_{-}=35$. The outer solutions require more care, as the boundary at infinity is an irregular singular point. Our expansion in Eq.~\eqref{eq:infinity_expansion} is an asymptotic series and, as such, the series is not strictly convergent in $n$ for a fixed $r_\infty$. The ansatz will initially show power law convergence with increasing $n$, but for sufficiently high $n$ the series will begin to diverge. At this point it is no longer useful to add higher order terms. Note for a fixed max value $n_+$ the series will still converge with increasing $r_\infty$ as expected. After analysing this behaviour, we take $n_{+}=100$ to get the best boundary conditions. Given the boundary expansions as a function of $r_{H/\infty}$, for fixed $n_{\pm}$, we must then choose a location for our boundary sufficiently close to $r_* =\pm \infty$ to give the desired accuracy. Setting the final term in our ansatz to be of order $10^{-d}$, where $d$ is our desired number of significant figures, we choose as our boundaries:
\begin{align}
	r_\infty & = (a_{n_+} 10^d)^{1/n_+}, \\
	r_H & = 2M+(b_{n_-} 10^d)^{-1/n_-} .
\end{align} 
The expansions \eqref{eq:horizon_expansion} and \eqref{eq:infinity_expansion} give the boundary conditions in terms of an arbitary overall amplitude, specified by $a_0$ and $b_0$. As we first construct homogeneous solutions we can set these amplitudes to any non-zero value, and in practice we choose $a_0=b_0=1$. The amplitudes are then fixed by the matching procedure described above.

\subsubsection{Numerical algorithm}

In this section we briefly outline the steps we take in our numerical calculation in the Regge-Wheeler gauge. The Lorenz-gauge calculation follows a very similar set of steps \cite{Akcay:2010}.
\begin{itemize}
	\item{For each $lm$-mode with $l\ge2$ solve the odd sector RW equation, even if $l+m=\text{even}$. For the radiative modes ($l\ge2,m\neq0$) calculate boundary conditions for the homogeneous fields at $r_{H/\infty}$ using Eqs.~\eqref{eq:horizon_expansion} and \eqref{eq:infinity_expansion}. Using the boundary conditions, numerically integrate the homogeneous field equation \eqref{eq:RW} from the boundaries to the particle's orbit at $r=r_0$. For the static modes ($l\ge2,m=0$) evaluate the static homogeneous solutions \eqref{eq:RW_static_odd_in}-\eqref{eq:RW_static_odd_out} at the particle. Store the values of the inner and outer homogeneous fields and their radial derivatives at $r_0$.}
	\item{For $l\ge2$ and $l+m=\text{even}$ transform from the odd sector homogeneous solutions to the even sector homogeneous solutions using Eq.~\eqref{eq:RW_to_Z}.}
	\item{For all modes with $l\ge2$ construct the inhomogeneous solutions via Eq.~\eqref{eq:inhom_sols}.}
	\item{For the $l\ge2$ modes reconstruct the metric perturbation using the formula in, e.g., Refs.~\cite{Berndtson:thesis}.}
	\item{Complete the metric perturbation using the monopole and dipole solutions given in Eqs.~\eqref{eq:RW_tt_monopole}-\eqref{eq:odd_dipole}.}
	\item{Compute the retarded field $l$-mode (summed over $m$) contributions to the octupolar invariants using the formulae in Appendix \ref{apdx:invariants_in_Schw_coords}. }
	\item{Construct the regularized $l$-modes using the the standard mode-sum approach. The resulting contributions to the mode-sum accumulate rather slowly as $l^{-2}$.}
	\item{Numerically fit for the unknown higher-order regularization parameters and use these to increase the rate of convergence of the mode-sum with $l$. This procedure is common in self-force calculations and is described in, e.g., Ref.~\cite{Akcay-Warburton-Barack:2013}. }
	\item{To get the final result sum over $l$ and make the shift to the asymptotically flat gauge as discussed in Appendix~\ref{sec:shift-asympt}.}
\end{itemize}

For $r_0\ge4M$ we set the maximum computed $l$-mode to be $l_\text{max}=80$. This is sufficient to compute the octupolar invariants to high accuracy -- see Sec.~\ref{sec:results} for details on the accuracy we obtain. For orbits with $3M<r_0<4M$ we find we need an increasing number of $l$-modes to achieve good accuracy in the final results, and for orbits near the light-ring (located at $r_0=3M$) we set $l_\text{max}=130$ in our code -- see Sec.~\ref{sec:light-ring_results}.

\subsection{Post-Newtonian expansion}\label{sec:PN_method}

The generation of analytic post-Newtonian expansions for the octupolar gauge invariants requires a calculation of the
homogeneous solutions of the Regge-Wheeler equation for each $\ell$ mode. 
A general strategy for doing this was described in \citep{Bini:2013rfa}. The calculation is broken into three sections: (i) the exact results of Zerilli give the 
$\ell=0,1$ components of the metric -- see Eqs.~\eqref{eq:RW_tt_monopole}-\eqref{eq:odd_dipole}, (ii) certain `low-$\ell$' values calculated using the series solutions of Mano, Suzuki and
Takasugi and (iii) `high-$\ell$' contributions using a post-Newtonian ansatz. In a recent paper \cite{Kavanagh:2015lva} this approach was optimised 
and improved allowing extremely high PN orders to be computed, which otherwise are only accessible by experimental mathematics techniques \cite{Johnson-McDaniel:2015vva,Shah:2013,Shah:2015nva,Shah:2014tka}. In the rest of this section we give a very brief overview of our technique and refer the reader to Ref.~\cite{Kavanagh:2015lva} for further details.

The analytic MST homogeneous solutions are expressed using an infinite series of hypergeometric functions denoted $X^{\text{in}}_{\ell m}$, which satisfies 
the required boundary conditions at the horizon, and a series of irregular confluent hypergeometric functions, $X^{\text{up}}_{\ell m}$, 
satisfying the boundary conditions as $r_{*}\rightarrow\infty$. Specifically we can write
\begin{align*}
X^{\text{in}}_{\ell m}&\sim B^{\text{trans}}_{\ell m}e^{+i \omega r_{*}} , \qquad r_{*}\rightarrow-\infty\\
X^{\text{up}}_{\ell m}&\sim C^{\text{trans}}_{\ell m}e^{-i \omega r_{*}} , \qquad r_{*}\rightarrow\infty
\end{align*}
so that, with $a_0=b_0=1$ in Eqs.~\eqref{eq:horizon_expansion} and \eqref{eq:infinity_expansion}, we have the identification
\begin{align}
X^{\text{in}}_{\ell m}&= B^{\text{trans}}_{\ell m} \tilde{\Psi}_{RW}^\text{H}  \\
X^{\text{up}}_{\ell m}&= C^{\text{trans}}_{\ell m} \tilde{\Psi}_{RW}^\infty .
\end{align}

For the purposes of doing a PN expansion of the solutions for a particle on
a circular orbit one finds two natural and related small parameters, the frequency $\omega$ and the inverse of the radius, which are 
related by $M \Omega=\sqrt{2 G M/r^3}$. A natural way to deal with this double expansion is to instead expand in $\eta=1/c$, and introduce
two auxiliary variables  $X_1=GM/r$, $X_2{}^{\!1/2}=\omega r$, so that each instance of $X_1$ and $X_2$ must each come with an $\eta^2$ 
and are of the same order in the large-$r$ limit. 
Expanding these solutions to a given PN order in this way amounts to truncating
the $X^{\text{in/up}}$ infinite series at a finite order. 
However an in depth analysis of the series coefficients and the sometimes subtle behaviour of the hypergeometric
functions reveals a structure that can be exploited to optimise this truncation order and fine tune the length of the expansion 
of each term in the series. 

A practical difficulty of this approach is that the MST series becomes increasingly large with higher $\eta$-order. For each
PN order $y\sim\frac{1}{r_\Omega}\sim \eta^2$ so that  to get say 10 PN, we need 20 $\eta$ powers. 
Significant further simplifications of these large series can be made rewriting the expansion as, for example,
\begin{align}
\label{eq:in_ansatz}
X^{\text{in(MST)}}_{\ell m}= e^{i \psi^\text{in}} X_1{}^{-\ell - 1 - 
 \sum\limits_{j=1}^{\infty} a_{(6 j, 2 j)} (2 X_1 X_2{}^{\!1/2} \eta^3)^{2 j}} \times \left[1+\eta^2 A_2^{\ell}+\eta^4 A_4^{\ell}+\eta^6 A_6^{\ell}+\ldots\right],
\end{align}
where the $A_i$ are \emph{strictly} polynomials in $X_1$, $X_2$. Since  $2 X_1 X_2{}^{\!1/2}=2 G M \omega$, we see that $\psi$ is $r$-independent allowing it to be essentially ignored 
as it will drop out with the wronskian during normalisation. We note that the purely even series in $\eta$ includes 
some odd powers that appear at $\ell$-dependent powers, and with these we also get extra unaccounted for log terms. For instance, for $\ell=2$ the first odd term is at $\eta^{13}$. 

As such, for a large-enough $\ell$ (dependent on the required expansion order), the homogeneous solutions become regular
enough to instead use an ansatz of purely even powers as the solution of the RW equation. The details of this are described thoroughly in 
\cite{Kavanagh:2015lva}. Once the homogeneous solutions of the Regge-Wheeler equation have been obtained, the even-parity solutions can be expressed using Eq.~\eqref{eq:RW_to_Z}. This allows us to reconstruct the full metric perturbation, and from there our gauge invariant quantities, entirely from the Regge-Wheeler series solutions. 

\section{Results}\label{sec:results}

In this section we present results for the octupolar invariants computed for circular orbits in a Schwarzschild background. More specifically, we present the six electric-type invariants defined in Eqs.~\eqref{Eijk:first}-\eqref{Eijk:last}, and the four magnetic-type invariants defined in Eqs.~\eqref{Bijk:first}-\eqref{Bijk:last}. % and the $\hat{\delta}_{K3+}$ defined in Eq.~\eqref{eq:hatdeltaK3p}.
In Sec.~\ref{subsec:data} we exhibit numerical data, and in Sec.~\ref{subsec:pn} we supply post-Newtonian expansions. In Sec.~\ref{sec:light-ring_results} we examine the behaviour of the invariants in the approach to the light-ring at $r=3M$. 

\subsection{Numerical data}\label{subsec:data}
\begin{figure}
	\includegraphics[width=8.5cm]{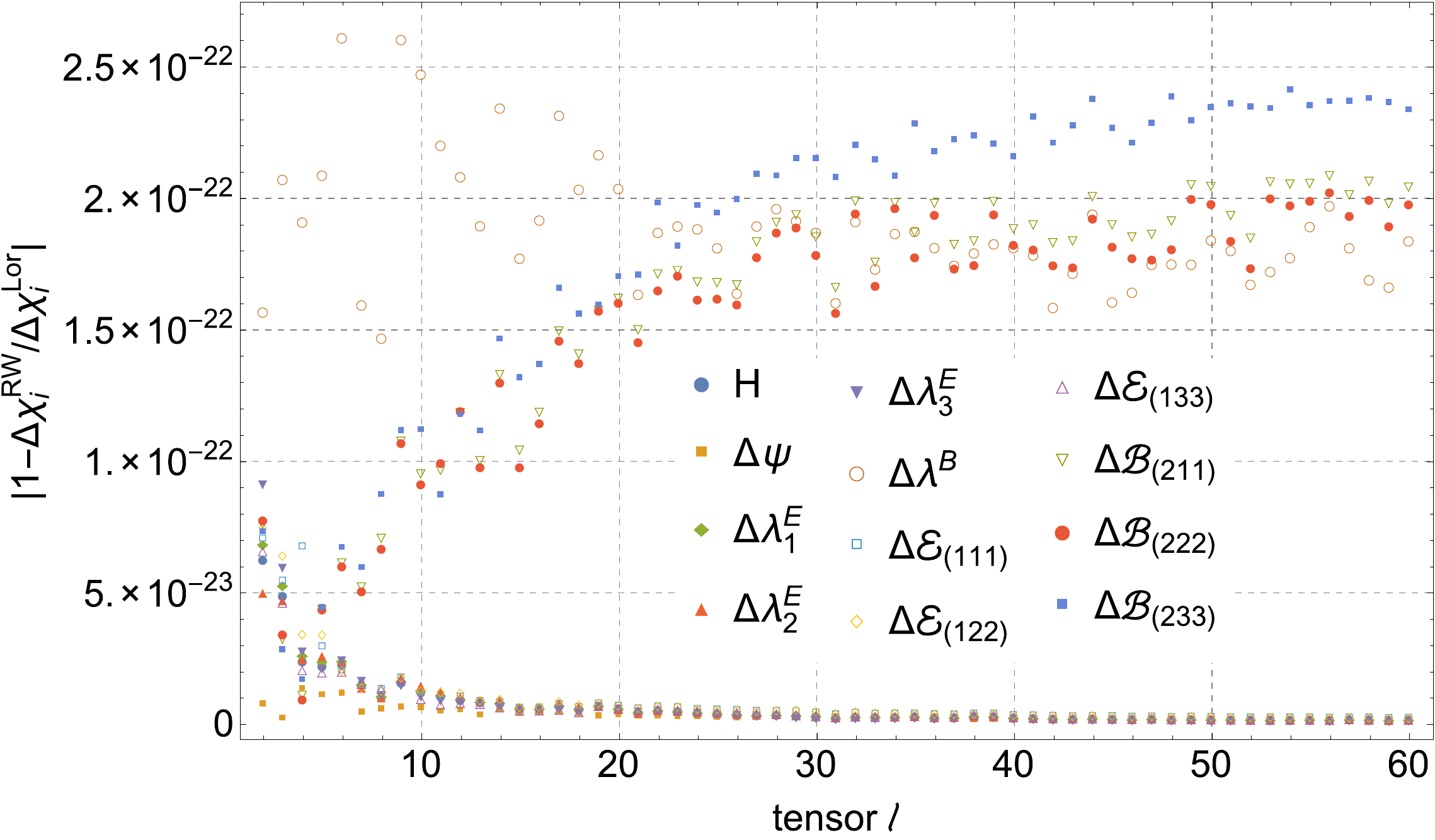}
	\caption{Comparison of numerical results computed in the RW and Lorenz gauges for a variety of conservative gauge invariant quantities, $\Delta\chi_i$, along a circular orbit at $r_0=10M$. We see 22--24 significant digits agreement in the individual tensor l-modes of the retarded field.}\label{fig:RW_vs_Lorenz}
\end{figure}

We have employed two independent calculations in the Regge-Wheeler and Lorenz gauges: see Sec.~\ref{sec:computational_approaches} or Ref.~\cite{Akcay:2010} for details, respectively. Both codes are implemented in Mathematica, which allows us to go beyond machine precision. We find that the Regge-Wheeler and Lorenz gauge results for retarded field contribution to the invariants agree to around 22--24 significant figures. This high level of agreement, exemplified in Fig.~\ref{fig:RW_vs_Lorenz},  increases our confidence in the validity of the numerical calculation.

In Table \ref{table:cons_electric} we present sample numerical results for the three conservative electric-type invariants. Table \ref{table:diss_electric} provides the results for the three dissipative electric-type invariants. As the computation of the latter does not involve a regularization step, the dissipative results are considerably more accurate than for the conservative results. Our numerical results for the three conservative and one dissipative magnetic-type invariants are presented in Table \ref{table:magnetic}. % and we present numerical results for $\hat{\delta}_{K3+}$ in Table \ref{table:dK3p}.

\begin{table}
\begin{center}
\begin{tabular}{c | c  c  c }
\hline\hline
$r_\Omega/M$ & $\Delta \mathcal{E}_{(111)}$ & $\Delta \mathcal{E}_{(122)}$ & $\Delta \mathcal{E}_{(133)}$ \\
\hline
$4$		& $-6.87640142\times 10^{-2}$		& $5.634572704\times 10^{-2}$		& $1.24182872\times 10^{-2}$	\\
$5$		& $-1.3622429846\times 10^{-2}$		& $9.45418747546\times 10^{-3}$		& $4.1682423703\times 10^{-3}$	\\
$6$		& $-5.61141083923\times 10^{-3}$	& $3.477232505498\times 10^{-3}$	& $2.13417833373\times 10^{-3}$	\\
$7$		& $-2.925643118454\times 10^{-3}$	& $1.701979164325\times 10^{-3}$	& $1.223663954129\times 10^{-3}$	\\
$8$		& $-1.710986615756\times 10^{-3}$	& $9.592475191788\times 10^{-4}$	& $7.517390965770\times 10^{-4}$	\\
$9$		& $-1.075500995896\times 10^{-3}$	& $5.890480037652\times 10^{-4}$	& $4.864529921306\times 10^{-4}$	\\
$10$	& $-7.120764484958\times 10^{-4}$	& $3.838876753995\times 10^{-4}$	& $3.281887730964\times 10^{-4}$	\\
$12$	& $-3.494517915911\times 10^{-4}$	& $1.847169590917\times 10^{-4}$	& $1.647348324994\times 10^{-4}$	\\
$14$	& $-1.913146810405\times 10^{-4}$	& $9.995537359206\times 10^{-5}$	& $9.135930744843\times 10^{-5}$	\\
$16$	& $-1.133949991793\times 10^{-4}$	& $5.879346730837\times 10^{-5}$	& $5.460153187097\times 10^{-5}$	\\
$18$	& $-7.141332604056\times 10^{-5}$	& $3.682750321689\times 10^{-5}$	& $3.458582282367\times 10^{-5}$	\\
$20$	& $-4.718352028785\times 10^{-5}$	& $2.423514347706\times 10^{-5}$	& $2.294837681079\times 10^{-5}$	\\
$30$	& $-9.514915883987\times 10^{-6}$	& $4.835793521499\times 10^{-6}$	& $4.679122362488\times 10^{-6}$	\\
$40$	& $-3.040712519124\times 10^{-6}$	& $1.538289606495\times 10^{-6}$	& $1.502422912629\times 10^{-6}$	\\
$50$	& $-1.252723439259\times 10^{-6}$	& $6.321181929625\times 10^{-7}$	& $6.206052462967\times 10^{-7}$	\\
$60$	& $-6.064208487551\times 10^{-7}$	& $3.054930741569\times 10^{-7}$	& $3.009277745982\times 10^{-7}$	\\
$70$	& $-3.282027079848\times 10^{-7}$	& $1.651475883984\times 10^{-7}$	& $1.630551195863\times 10^{-7}$	\\
$80$	& $-1.927657419028\times 10^{-7}$	& $9.691577786310\times 10^{-8}$	& $9.584996403967\times 10^{-8}$	\\
$90$	& $-1.205253640043\times 10^{-7}$	& $6.055682885677\times 10^{-8}$	& $5.996853514753\times 10^{-8}$	\\
$100$	& $-7.917190975864\times 10^{-8}$	& $3.975890910078\times 10^{-8}$	& $3.941300065787\times 10^{-8}$	\\
$500$	& $-1.277421047615\times 10^{-10}$	& $6.392477321681\times 10^{-11}$	& $6.381733154472\times 10^{-11}$	\\
$1000$	& $-7.991970194046\times 10^{-12}$	& $3.997657790884\times 10^{-12}$	& $3.994312403162\times 10^{-12}$	\\
$5000$	& $-1.279743808249\times 10^{-14}$	& $6.399252758926\times 10^{-15}$	& $6.398185323566\times 10^{-15}$	\\
\hline
\end{tabular}
\end{center}
\caption{Sample numerical results for the conservative electric-type octupolar invariants.}\label{table:cons_electric}
\end{table}

\begin{table}
\begin{center}
\begin{tabular}{c | c  c  c }
\hline\hline
$r_\Omega/M$ & $\Delta \mathcal{E}_{(113)}$ & $\Delta \mathcal{E}_{(223)}$ & $\Delta \mathcal{E}_{(333)}$ \\
\hline
$4$	& $1.43018712098924\times 10^{-2}$	& $-6.81363125080514\times 10^{-3}$	&$-7.48823995908726\times 10^{-3}$	\\
$5$	& $1.69051912392376\times 10^{-3}$	& $-6.68228419170062\times 10^{-4}$	&$-1.02229070475370\times 10^{-3}$	\\
$6$	& $3.93615041880796\times 10^{-4}$	& $-1.40326772303052\times 10^{-4}$	&$-2.53288269577744\times 10^{-4}$	\\
$7$	& $1.24851076918558\times 10^{-4}$	& $-4.16707182592131\times 10^{-5}$	&$-8.31803586593453\times 10^{-5}$	\\
$8$	& $4.78575862364605\times 10^{-5}$	& $-1.52593581419753\times 10^{-5}$	&$-3.25982280944852\times 10^{-5}$	\\
$9$	& $2.09252624095044\times 10^{-5}$	& $-6.45207930480653\times 10^{-6}$	&$-1.44731831046978\times 10^{-5}$	\\
$10$	& $1.00921765694192\times 10^{-5}$	& $-3.03317966765418\times 10^{-6}$	&$-7.05899690176504\times 10^{-6}$	\\
$12$	& $2.90853534249746\times 10^{-6}$	& $-8.42878520552755\times 10^{-7}$	&$-2.06565682194470\times 10^{-6}$	\\
$14$	& $1.02905687355232\times 10^{-6}$	& $-2.90962875240999\times 10^{-7}$	&$-7.38093998311317\times 10^{-7}$	\\
$16$	& $4.21307269886127\times 10^{-7}$	& $-1.17032511794287\times 10^{-7}$	&$-3.04274758091840\times 10^{-7}$	\\
$18$	& $1.92451417988312\times 10^{-7}$	& $-5.27522804430828\times 10^{-8}$	&$-1.39699137545229\times 10^{-7}$	\\
$20$	& $9.57423553217574\times 10^{-8}$	& $-2.59726822309717\times 10^{-8}$	&$-6.97696730907857\times 10^{-8}$	\\
$30$	& $6.63075048503344\times 10^{-9}$	& $-1.74627612621227\times 10^{-9}$	&$-4.88447435882117\times 10^{-9}$	\\
$40$	& $1.00806706036123\times 10^{-9}$	& $-2.61810134057605\times 10^{-10}$	&$-7.46256926303620\times 10^{-10}$	\\
$50$	& $2.34720446527898\times 10^{-10}$	& $-6.04700459200130\times 10^{-11}$	&$-1.74250400607885\times 10^{-10}$	\\
$60$	& $7.14698583248068\times 10^{-11}$	& $-1.83158348544703\times 10^{-11}$	&$-5.31540234703365\times 10^{-11}$	\\
$70$	& $2.61736513652863\times 10^{-11}$	& $-6.68285229858544\times 10^{-12}$	&$-1.94907990667008\times 10^{-11}$	\\
$80$	& $1.09692221205352\times 10^{-11}$	& $-2.79307929655166\times 10^{-12}$	&$-8.17614282398351\times 10^{-12}$	\\
$90$	& $5.09523969822615\times 10^{-12}$	& $-1.29465822904663\times 10^{-12}$	&$-3.80058146917952\times 10^{-12}$	\\
$100$	& $2.56663032262918\times 10^{-12}$	& $-6.51067248226772\times 10^{-13}$	&$-1.91556307440241\times 10^{-12}$	\\
$500$	& $7.32252735609857\times 10^{-17}$	& $-1.83582572460285\times 10^{-17}$	&$-5.48670163149571\times 10^{-17}$	\\
$1000$	& $8.09167955607435\times 10^{-19}$	& $-2.02577720909791\times 10^{-19}$	&$-6.06590234697644\times 10^{-19}$	\\
$5000$	& $2.31673725041822\times 10^{-23}$	& $-5.79347353907946\times 10^{-24}$	&$-1.73738989651028\times 10^{-23}$	\\
\hline
\end{tabular}
\end{center}
\caption{Sample numerical results for the dissipative electric-type octupolar invariants.}\label{table:diss_electric}
\end{table}

\begin{table}
\begin{center}
\begin{tabular}{c | c  c  c c }
\hline\hline
$r_\Omega/M$ & $\Delta \mathcal{B}_{(211)}$ & $\Delta \mathcal{B}_{(222)}$ & $\Delta \mathcal{B}_{(233)}$ & $\Delta \mathcal{B}_{(123)}$ \\
\hline
$4$	& $-6.148298254370\times 10^{-2}$	&$5.070286329453\times 10^{-2}$	&$1.078011924917\times 10^{-2}$	&$1.07801192491724\times 10^{-2}$	\\
$5$	& $-9.558323357929\times 10^{-3}$	&$7.670992694990\times 10^{-3}$	&$1.887330662938\times 10^{-3}$	&$1.88733066293848\times 10^{-3}$	\\
$6$	& $-3.155936380263\times 10^{-3}$	&$2.476758241817\times 10^{-3}$	&$6.791781384454\times 10^{-4}$	&$6.79178138445361\times 10^{-4}$	\\
$7$	& $-1.397948966284\times 10^{-3}$	&$1.081923065789\times 10^{-3}$	&$3.160259004949\times 10^{-4}$	&$3.16025900494923\times 10^{-4}$	\\
$8$	& $-7.234703923371\times 10^{-4}$	&$5.550936330078\times 10^{-4}$	&$1.683767593293\times 10^{-4}$	&$1.68376759329306\times 10^{-4}$	\\
$9$	& $-4.130973443372\times 10^{-4}$	&$3.151840931693\times 10^{-4}$	&$9.791325116795\times 10^{-5}$	&$9.79132511679517\times 10^{-5}$	\\
$10$	& $-2.528015715619\times 10^{-4}$	&$1.921517184307\times 10^{-4}$	&$6.064985313116\times 10^{-5}$	&$6.06498531311616\times 10^{-5}$	\\
$12$	& $-1.095551773983\times 10^{-4}$	&$8.288773695651\times 10^{-5}$	&$2.666744044177\times 10^{-5}$	&$2.66674404417714\times 10^{-5}$	\\
$14$	& $-5.444485917231\times 10^{-5}$	&$4.108569884562\times 10^{-5}$	&$1.335916032669\times 10^{-5}$	&$1.33591603266878\times 10^{-5}$	\\
$16$	& $-2.980264003325\times 10^{-5}$	&$2.245424872606\times 10^{-5}$	&$7.348391307187\times 10^{-6}$	&$7.34839130718667\times 10^{-6}$	\\
$18$	& $-1.753993694654\times 10^{-5}$	&$1.320134773357\times 10^{-5}$	&$4.338589212967\times 10^{-6}$	&$4.33858921296681\times 10^{-6}$	\\
$20$	& $-1.092394842331\times 10^{-5}$	&$8.215915676267\times 10^{-6}$	&$2.708032747040\times 10^{-6}$	&$2.70803274704046\times 10^{-6}$	\\
$30$	& $-1.770249199828\times 10^{-6}$	&$1.329249843091\times 10^{-6}$	&$4.409993567363\times 10^{-7}$	&$4.40999356736253\times 10^{-7}$	\\
$40$	& $-4.868817160862\times 10^{-7}$	&$3.653967138885\times 10^{-7}$	&$1.214850021976\times 10^{-7}$	&$1.21485002197624\times 10^{-7}$	\\
$50$	& $-1.788310960062\times 10^{-7}$	&$1.341778109443\times 10^{-7}$	&$4.465328506186\times 10^{-8}$	&$4.46532850618555\times 10^{-8}$	\\
$60$	& $-7.887212354333\times 10^{-8}$	&$5.917061308384\times 10^{-8}$	&$1.970151045949\times 10^{-8}$	&$1.97015104594935\times 10^{-8}$	\\
$70$	& $-3.946891664217\times 10^{-8}$	&$2.960771807198\times 10^{-8}$	&$9.861198570192\times 10^{-9}$	&$9.86119857019213\times 10^{-9}$	\\
$80$	& $-2.166444813367\times 10^{-8}$	&$1.625085708368\times 10^{-8}$	&$5.413591049991\times 10^{-9}$	&$5.41359104999132\times 10^{-9}$	\\
$90$	& $-1.276212037585\times 10^{-8}$	&$9.572758888543\times 10^{-9}$	&$3.189361487310\times 10^{-9}$	&$3.18936148731046\times 10^{-9}$	\\
$100$	& $-7.948907544564\times 10^{-9}$	&$5.962268324527\times 10^{-9}$	&$1.986639220037\times 10^{-9}$	&$1.98663922003664\times 10^{-9}$	\\
$500$	& $-5.716760192009\times 10^{-12}$	&$4.287586670256\times 10^{-12}$	&$1.429173521752\times 10^{-12}$	&$1.42917352175245\times 10^{-12}$	\\
$1000$	& $-2.528141980419\times 10^{-13}$	&$1.896108307399\times 10^{-13}$	&$6.320336730204\times 10^{-14}$	&$6.32033673020413\times 10^{-14}$	\\
$5000$	& $-1.809952182177\times 10^{-16}$	&$1.357464188697\times 10^{-16}$	&$4.524879934796\times 10^{-17}$	&$4.52487993479633\times 10^{-17}$	\\
\hline
\end{tabular}
\end{center}
\caption{Sample numerical results for the magnetic-type octupolar invariants.}\label{table:magnetic}
\end{table}

\subsection{Post-Newtonian expansions}\label{subsec:pn}
As outlined in Sec.~\ref{sec:PN_method}, we have made a post-Newtonian calculation of the octupolar invariants using a method which builds upon the work of Ref.~\cite{Kavanagh:2015lva}. This method allows us to take the expansions to very high order.  Results at 15th post-Newtonian order are available in an online repository \cite{online}. Here, for brevity, we truncate the displayed results at a relatively low order:
\begin{eqnarray}
 \Delta \EE_{(111)} &=& -8 y^4 + 8 y^5 + 30 y^6 - (\tfrac{1711}{6} - \tfrac{4681}{512} \pi^2) y^7 + \bigl(\tfrac{136099}{400} -  \tfrac{6255}{1024} \pi^2 -  \tfrac{2048}{5} \gamma -  \tfrac{4096}{5} \log 2 -  \tfrac{1024}{5} \log y\bigr) y^8 \nonumber \\
&& - \bigl(\tfrac{1604627}{630} - \tfrac{6413231}{49152} \pi^2 - \tfrac{159664}{105} \gamma - \tfrac{18416}{5} \log 2 + \tfrac{4374}{7} \log 3 - \tfrac{79832}{105} \log y\bigr) y^9 -  \tfrac{219136}{525} \pi y^{19/2} + \mathcal{O}\bigl(y^{10}\bigr),
\\
 \Delta \EE_{(122)} &=& 4 y^4 -  \tfrac{7}{3} y^5 - 9 y^6 + (\tfrac{1369}{8} -  \tfrac{9677}{2048} \pi^2) y^7 + \bigl(\tfrac{121369}{7200} + \tfrac{265}{192} \pi^2 + \tfrac{1024}{5} \gamma + \tfrac{2048}{5} \log 2 + \tfrac{512}{5} \log y\bigr) y^8 \nonumber \\
&& - \bigl(\tfrac{132611239}{120960} - \tfrac{240298535}{1179648} \pi^2 + \tfrac{173416}{315} \gamma + \tfrac{53416}{35} \log 2 - \tfrac{2916}{7} \log 3 + \tfrac{86708}{315} \log y\bigr) y^9 + \tfrac{109568}{525} \pi y^{19/2} + \mathcal{O}\bigl(y^{10}\bigr), \nonumber \\
\\
 \Delta \EE_{(133)} &=& 4 y^4 -  \tfrac{17}{3} y^5 - 21 y^6 + (\tfrac{2737}{24} -  \tfrac{9047}{2048} \pi^2) y^7 - \bigl(\tfrac{2571151}{7200} - \tfrac{14525}{3072} \pi^2 - \tfrac{1024}{5} \gamma - \tfrac{2048}{5} \log 2 - \tfrac{512}{5} \log y\bigr) y^8 \nonumber \\
&& + \bigl(\tfrac{62957089}{17280} -  \tfrac{394216079}{1179648} \pi^2 -  \tfrac{305576}{315} \gamma -  \tfrac{75496}{35} \log 2 + \tfrac{1458}{7} \log 3 -  \tfrac{152788}{315} \log y\bigr) y^9 + \tfrac{109568}{525} \pi y^{19/2} + \mathcal{O}\bigl(y^{10}\bigr), \nonumber \\
\\
 \Delta \EE_{(113)} &=& \tfrac{128}{5} y^{13/2} -  \tfrac{108}{5} y^{15/2} + \tfrac{512}{5} \pi y^8 -  \tfrac{46978}{105} y^{17/2} + \tfrac{3794}{45} \pi y^9 \nonumber \\
&& + \tfrac{8}{496125} \bigl(107554351 + 8467200 \pi^2 - 25885440 \gamma - 51770880 \log 2 - 12942720 \log y\bigr) y^{19/2} + \mathcal{O}\bigl(y^{10}\bigr),
\\
 \Delta \EE_{(223)} &=& - \tfrac{32}{5} y^{13/2} -  \tfrac{18}{5} y^{15/2} -  \tfrac{128}{5} \pi y^8 + \tfrac{8276}{105} y^{17/2} -  \tfrac{15242}{315} \pi y^9 \nonumber \\
&& - \tfrac{1}{496125} \bigl(152535527 + 16934400 \pi^2 - 51770880 \gamma - 103541760 \log 2 - 25885440 \log y\bigr) y^{19/2} + \mathcal{O}\bigl(y^{10}\bigr),
\\
 \Delta \EE_{(333)} &=& - \tfrac{96}{5} y^{13/2} + \tfrac{126}{5} y^{15/2} -  \tfrac{384}{5} \pi y^8 + \tfrac{38702}{105} y^{17/2} -  \tfrac{3772}{105} \pi y^9 \nonumber \\
&& - \tfrac{1}{165375} \bigl(235966427 + 16934400 \pi^2 - 51770880 \gamma - 103541760 \log 2 - 25885440 \log y\bigr) y^{19/2} + \mathcal{O}\bigl(y^{10}\bigr),
\\
 \Delta \BB_{(123)} &=& \tfrac{64}{3} y^7 + \tfrac{36}{5} y^8 + \tfrac{256}{3} \pi y^{17/2} -  \tfrac{5347}{15} y^9 + \tfrac{2197}{15} \pi y^{19/2} \nonumber \\
&& + \tfrac{4}{11025} \bigl(3475113 + 313600 \pi^2 - 958720 \gamma - 1917440 \log 2 - 479360 \log y\bigr) y^{10} -  \tfrac{10961}{7} \pi y^{21/2} + \mathcal{O}\bigl(y^{11}\bigr),
\\
 \Delta \BB_{(211)} &=& -8 y^{9/2} + \tfrac{16}{3} y^{11/2} - 20 y^{13/2} + (- \tfrac{677}{2} + \tfrac{5101}{512} \pi^2) y^{15/2} \nonumber \\
&& - \tfrac{1}{230400} \bigl(246270016 - 20642025 \pi^2 + 94371840 \gamma + 188743680 \log 2 + 47185920 \log y\bigr) y^{17/2} \nonumber \\
&& - \tfrac{1}{154828800} \bigl(417740314624 - 17848070625 \pi^2 - 131939696640 \gamma - 390644367360 \log 2 \nonumber \\
&& \qquad + 123619737600 \log 3 - 65969848320 \log y\bigr) y^{19/2} -  \tfrac{219136}{525} \pi y^{10} + \mathcal{O}\bigl(y^{21/2}\bigr),
\\
 \Delta \BB_{(222)} &=& 6 y^{9/2} - 4 y^{11/2} + \tfrac{83}{4} y^{13/2} + (\tfrac{1069}{4} -  \tfrac{7809}{1024} \pi^2) y^{15/2} \nonumber \\
&& + \tfrac{1}{204800} \bigl(234195584 - 19194125 \pi^2 + 62914560 \gamma + 125829120 \log 2 + 31457280 \log y\bigr) y^{17/2} \nonumber \\
&& + \tfrac{1}{11468800} \bigl(125170823168 - 11193257425 \pi^2 - 5923143680 \gamma - 19177144320 \log 2 \nonumber \\
&& \qquad + 7166361600 \log 3 - 2961571840 \log y\bigr) y^{19/2} + \tfrac{54784}{175} \pi y^{10} + \mathcal{O}\bigl(y^{21/2}\bigr),
\\
 \Delta \BB_{(233)} &=& 2 y^{9/2} -  \tfrac{4}{3} y^{11/2} -  \tfrac{3}{4} y^{13/2} + (\tfrac{285}{4} -  \tfrac{2393}{1024} \pi^2) y^{15/2} \nonumber \\
&& - \tfrac{1}{1843200} \bigl(137600128 - 7610925 \pi^2 - 188743680 \gamma - 377487360 \log 2 - 94371840 \log y\bigr) y^{17/2} \nonumber \\
&& - \tfrac{1}{309657600} \bigl(2544131596288 - 266521809225 \pi^2 + 103954513920 \gamma + 263505838080 \log 2 \nonumber \\
&& \qquad - 53747712000 \log 3 + 51977256960 \log y\bigr) y^{19/2} + \tfrac{54784}{525} \pi y^{10} + \mathcal{O}\bigl(y^{21/2}\bigr).
\end{eqnarray}
where here $y=M/r_0$. 

\begin{figure}
	\includegraphics[width=8.5cm]{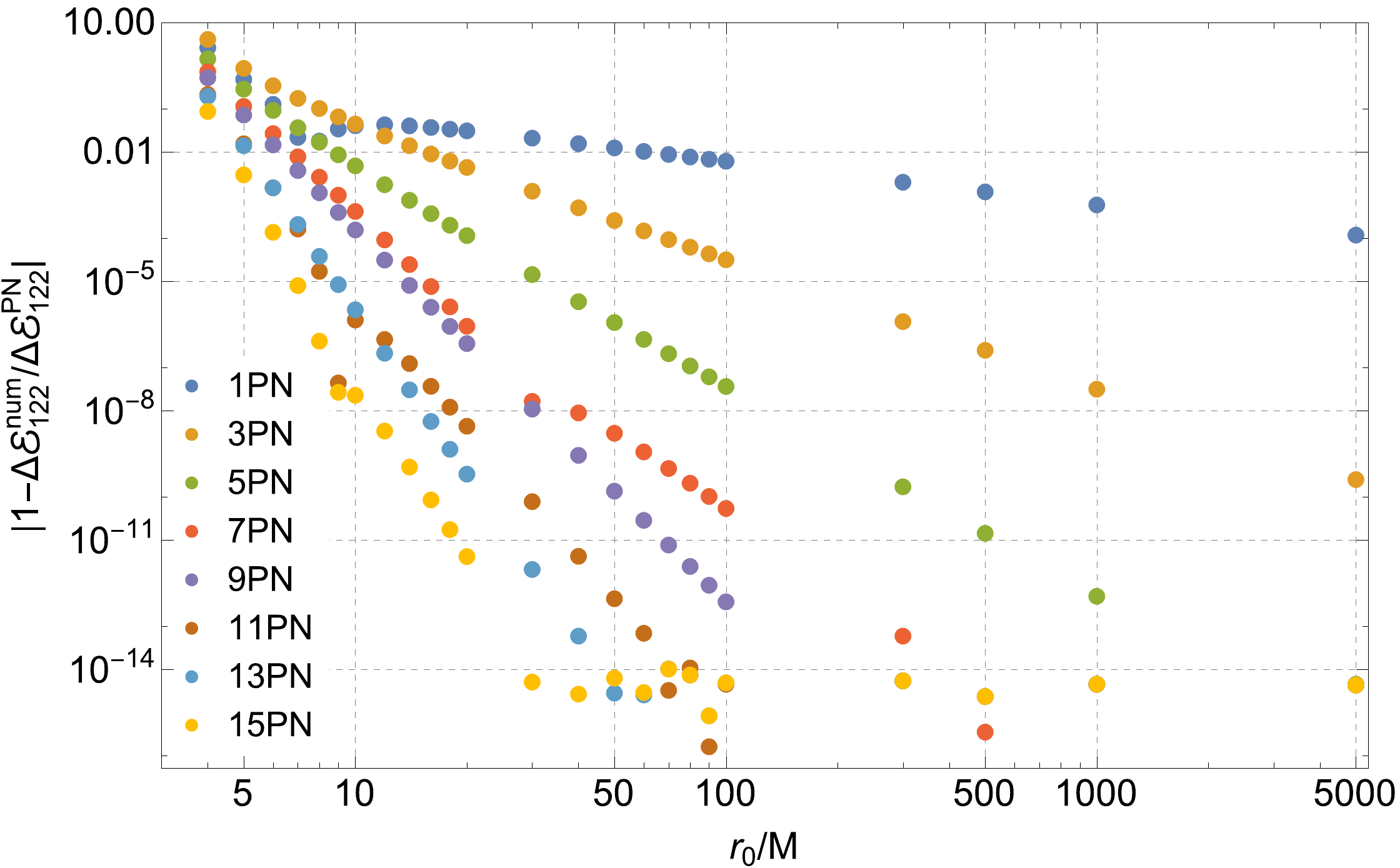}\hskip0.5cm
	\includegraphics[width=8.8cm]{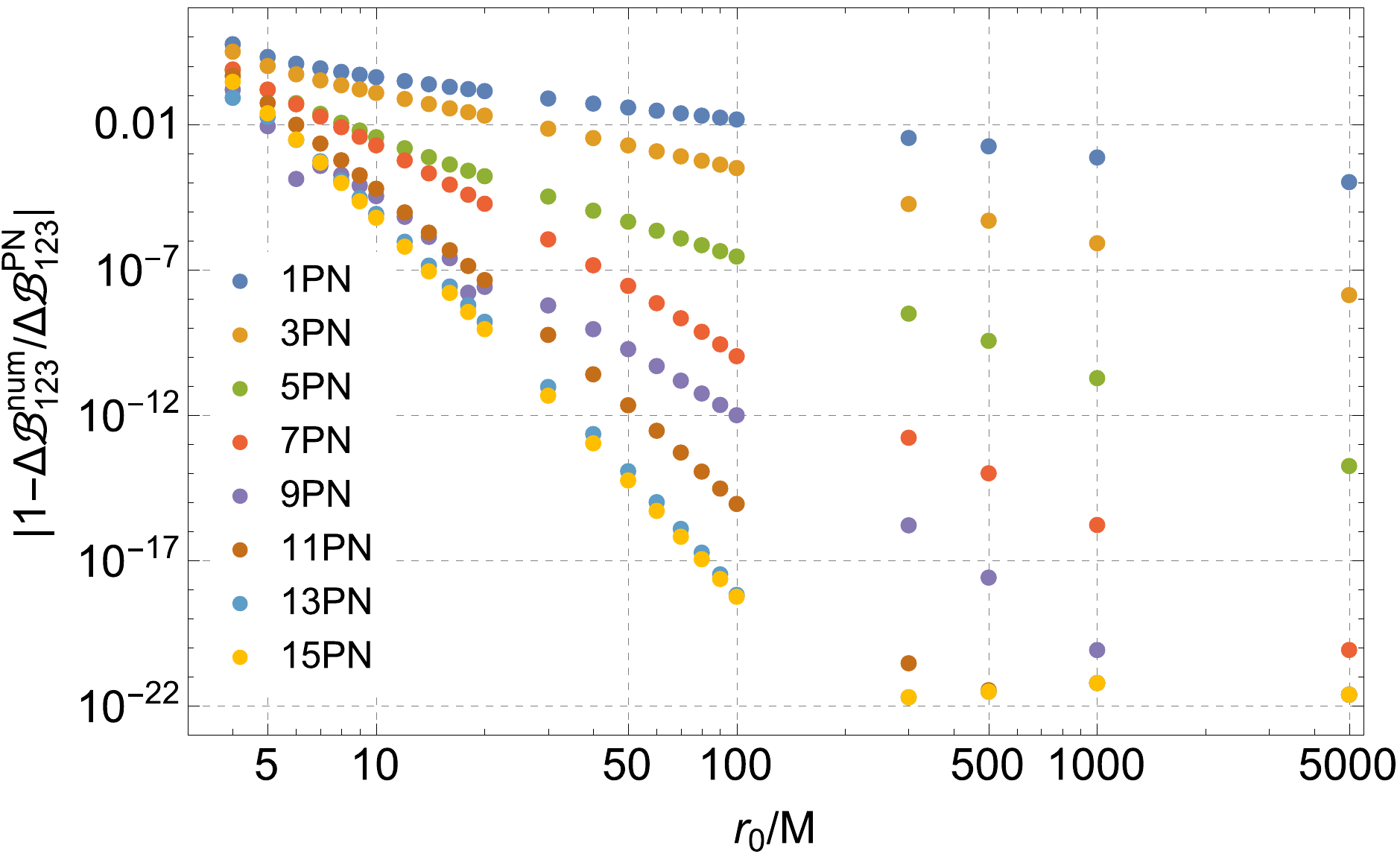}
	\caption{Comparison of our numerical and PN results for (left) $\Delta\mathcal{E}_{122}$ and (right) $\Delta\mathcal{B}_{123}$. For each invariant we plot the relative difference between the numerical data and successive truncations of the relevant PN series, i.e., in the legend `$x$PN' means we are comparing against the PN series with all terms up to and including (relative) $x$PN order. As successive PN terms are added the agreement between the PN series and the numerical results improves. For the conservative invariants, such as $\Delta\mathcal{E}_{122}$, the agreement between the PN series and the numerical data saturates at a relative accuracy of 13--14 significant figures. For the dissipative invariants, such as $\Delta\mathcal{B}_{123}$, the comparison saturates at 21--22 significant figures. This difference in accuracy in the numerical data stems from the requirement to regularize the conservative invariants whereas the dissipative invariants do not require regularization. }\label{fig:num_vs_PN}
\end{figure}

Figure \ref{fig:num_vs_PN} shows sample comparisons of our PN and numerical results. We observe that, as higher-order PN terms are included in the comparison, the agreement improves for all values of $r_0$. For large orbital radii the comparison saturates at the level of our (smaller than machine precision) numerical round-off error. For strong-field orbits, the comparison allows us to estimate how well the PN series performs in this regime. At $r_0=10M$ we typically find that the 15PN series recovers the first 7--8 significant digits of the numerical result. At the innermost stable circular orbit, at $r_0=6M$, the 15PN series successfully recovers the first 3--4 significant figures. The excellent agreement we observe between our PN and numerical calculations gives us further confidence in both sets of results.

% Include some nice comparison plots

\subsection{Behaviour near the light-ring}\label{sec:light-ring_results}

With our numerical codes we can calculate the behaviour of the octupolar invariants as the orbit approaches the light-ring at $r_0=3M$. In general, the invariants will diverge as the light-ring is approached, and knowledge of the rate of divergence, along with our high-order PN results and our other numerical results, may be useful in performing global fits for the invariants across all orbital radii. Such fits find utility in EOB theory and already results for the redshift, spin precession and tidal invariants have been employed in EOB models \cite{Akcay:2012ea,Bini:2014ica,Bini:2014zxa}. In this section we discuss, and give results for, the rate of divergence of the invariants near the light-ring but stop short of making global fits for the invariants.

The main challenge in computing conservative invariants near the light-ring is the late onset of convergence of the mode-sum in this regime (see Ref.~\cite{Akcay:2012ea} for a discussion of this behaviour). This necessitates computing a great deal more $lm$-modes; typically we set $l_\text{max}=130$ for our calculations in this regime. By comparison, for orbits with $r_0=4M$ we use $l_\text{max}=80$. Not only then do we need to numerically compute an additional 8085 $lm$-modes, on top of the 3239 modes required to reach $l_\text{max}=80$, but these higher $lm$-modes are more challenging to calculate numerically owing to the stronger power-law growth near the particle for high $l$ and the high mode frequency (and thus large number of oscillations that need to be resolved far from the particle) for high $m$-modes. These considerations mean that numerical calculations at radii near the light-ring are substantially more computationally expensive than our other numerical results.

\begin{figure}
	\includegraphics[width=8.5cm]{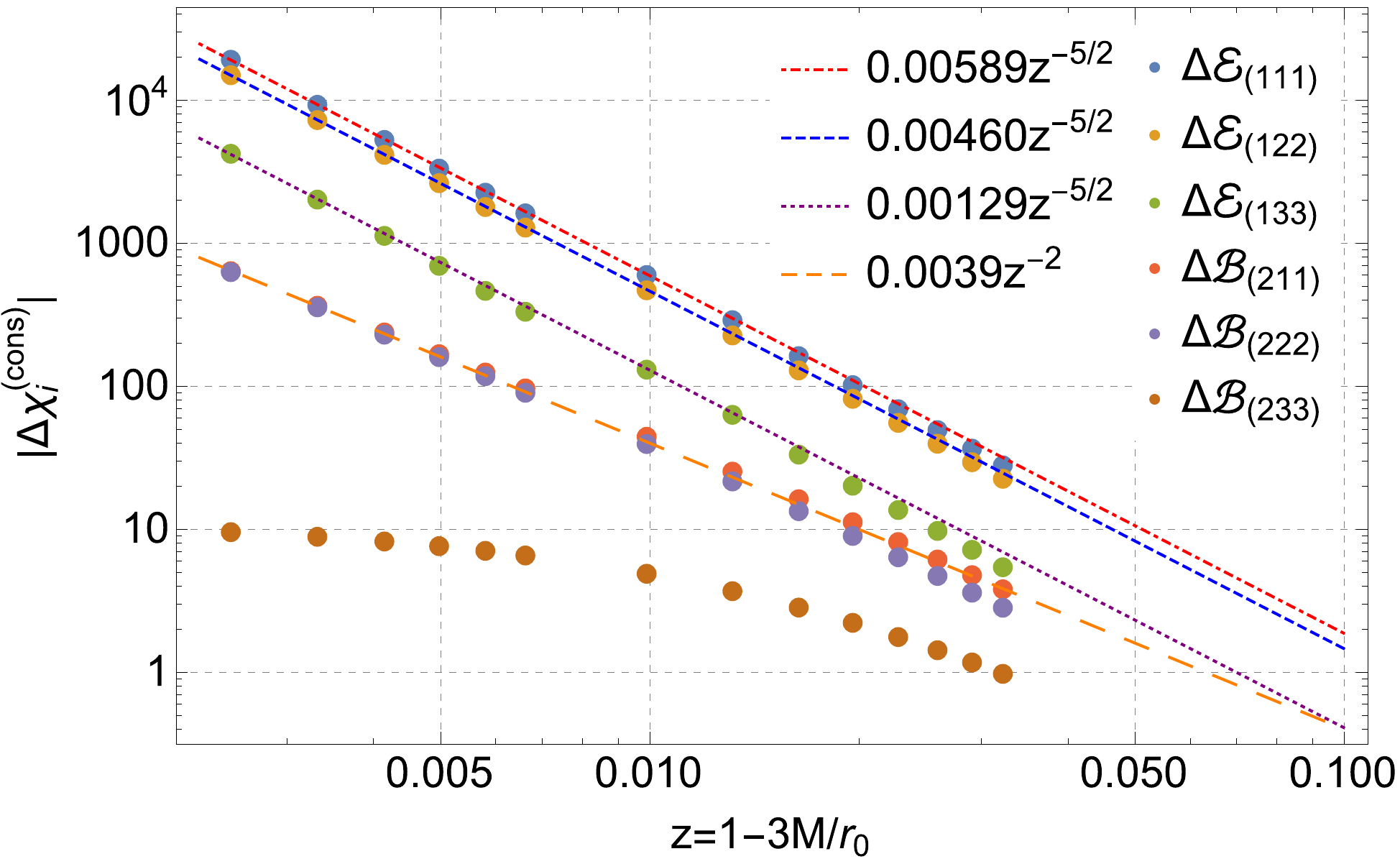}
	\caption{Divergence of the conservative octupolar invariants as the orbital radius approaches the light-ring. The electric-type invariants, $\Delta\mathcal{E}_{(111)},\Delta\mathcal{E}_{(122)},\Delta\mathcal{E}_{(133)}$, diverge as $z^{-5/2}$ where $z=1-3M/r_0$. Two of the magnetic-type invariants, $\Delta\mathcal{B}_{(211)}$ and $\Delta\mathcal{B}_{(222)}$, are observed to diverge as $z^{-2}$. We are unable to accurately deduce the rate of the divergence $\Delta\mathcal{B}_{(233)}$ but we plot our numerical results to show that its rate of divergence is subdominant to the other invariants.}\label{fig:light-ring}
\end{figure}

Our main results are presented in Fig.~\ref{fig:light-ring}. We are able to infer the rate of divergence of five out of six of the electric- and magnetic-type invariants. Defining $z\equiv1-3M/r_0$ we find $\Delta\mathcal{E}_{(111)}\sim-0.00589z^{-5/2}$, $\Delta\mathcal{E}_{(122)}\sim0.00406z^{-5/2}$, $\Delta\mathcal{E}_{(133)}\sim0.0129z^{-5/2}$, $\mathcal{B}_{(211)}\sim-0.0039z^{-2}$ and $\mathcal{B}_{(222)}\sim0.0039z^{-2}$ as $z\rightarrow0$. For the remaining conservative invariant, $\Delta\mathcal{B}_{(233)}$, our current results are not sufficient to accurately determine the divergence rate, but we can say that the rate is subdominant to the other invariants.

\section{Applications}\label{sec:applications}

Here we briefly outline two possible applications of the results of Sec.~\ref{sec:results}: in informing EOB theory, and in refining initial data for binary black hole simulations with large mass ratios in the strong field.

\subsection{Informing EOB theory}
In EOB theory, the dynamics of binary systems are reformulated in terms of the dynamics of a single ``effective'' body moving in a metric $ds^2 = -A(u;\nu)dt^2 + B(u;\nu)d\hat{r}^2 + \hat{r}^2\left(d\theta^2 + \sin^2\theta d\phi^2\right)$ (non-spinning case), where $A(u;\nu)$ and $B(u;\nu)$ are smooth functions of inverse radius $u = (M+\mu)/\hat{r}$ and symmetric mass ratio $\nu = \mu M / (\mu+M)^2$.  For tidal interactions, it was proposed in Ref.~\cite{Bini:2012gu} that the metric function should take the form $A = A^\text{BBH} + A_1^{\text{tidal}} + A_2^{\text{tidal}}$. The latter terms are radial potentials associated with tidal deformations of bodies $1$ and $2$, which may be decomposed into multipolar contributions, $A_1^{\text{tidal}} = A_1^{(2+)} + A_1^{(2-)} + A_1^{(3+)} + A_1^{(3-)} \ldots$, from the electric quadrupole ($A_1^{(2+)}$), magnetic quadrupole ($A_1^{(2-)}$), electric octupole ($A_1^{(3+)}$), magnetic octupole ($A_1^{(3-)}$) sectors, respectively, etc. In Ref.~\cite{Bini:2012gu} a relationship was established between the \emph{dynamically}-significant tidal functions $A_i^{(j\pm)}$ and \emph{kinematically}-invariant functions $J_{\bullet}(y)$ formed from the tidal tensors (see Eq.~(6.11) in Ref.~\cite{Bini:2014zxa}). In the quadrupolar sector, the relevant invariants are 
\beq
J_{e^2} \equiv \EE_{ab} \EE^{ab} , \quad J_{b^2} \equiv \BB_{ab} \BB^{ab} , \quad
J_{e^3} \equiv \EE_{ab} \EE^{bc} \EE^a_c,  \quad \ldots
\eeq
In the electric-octupolar sector, the relevant quantities are (see Appendix D of \cite{Bini:2014zxa}) $J_{3+} = K_{3+} + \frac{1}{3} J_{\dot{2}+}$, where $K_{3+} \equiv \EE_{(abc)}\EE^{(abc)}$ and $J_{\dot{2}+} \equiv \EE_{ab0} \EE^{ab0}$. In the magnetic-octupolar sector, analogous quantities may be formed.

The $\mathcal{O}(\mu)$ part of these invariants may be easily deduced from our octupolar components $\Delta \EE_{111}, \ldots$. For example, $\Delta K_{3+}$, obtained via Eq.~(\ref{eq:K3p-def}), is related to $\hat{\delta}_{K3+}$ by (\ref{eq:hatdeltaK3p}). 

Previously, Bini \& Damour have given a PN expansion of $\hat{\delta}_{K3+}$ to 7.5PN order (see Eq.~(D10) in Ref.~\cite{Bini:2014zxa}). With the results of Sec.~\ref{sec:results}, we are able to go a step further. First, in Table \ref{table:dK3p} we give numerical data for $\hat{\delta}_{K3+}$ in the strong-field regime. The data indicates that $\hat{\delta}_{K3+}$ has a local maximum somewhat within the innermost stable circular orbit. Second, in an online repository \cite{online}, we provide a higher-order PN expansion of $\hat{\delta}_{K3+}$; below, we state the expansion at 8.5PN order (correcting a very minor error/typo in the $y^6$ term of (D10) in Ref.~\cite{Bini:2014zxa}):
\begin{eqnarray}
  \hat{\delta}_{K3+} &=& - \tfrac{8}{3} + \tfrac{358}{45} y + \tfrac{11848}{675} y^2 + (- \tfrac{3581903}{40500} + \tfrac{4681}{1536} \pi^2) y^3 + \bigl(\tfrac{614794483}{2430000} -  \tfrac{790931}{92160} \pi^2 -  \tfrac{2048}{15} \gamma -  \tfrac{4096}{15} \log 2 -  \tfrac{1024}{15} \log y\bigr) y^4 \nonumber \\ 
&& + \bigl(- \tfrac{759123028241}{1020600000} + \tfrac{431520437}{11059200} \pi^2 + \tfrac{1070704}{1575} \gamma + \tfrac{354064}{225} \log 2 -  \tfrac{1458}{7} \log 3 + \tfrac{535352}{1575} \log y\bigr) y^5 -  \tfrac{219136}{1575} \pi y^{11/2} \nonumber \\ 
&& + \bigl(\tfrac{12569905047667}{2187000000} -  \tfrac{1903269674027}{1769472000} \pi^2 -  \tfrac{42147341}{6291456} \pi^4 + \tfrac{181080056}{212625} \gamma -  \tfrac{123628168}{212625} \log 2 + \tfrac{73953}{35} \log 3 + \tfrac{90540028}{212625} \log y\bigr) y^6 \nonumber \\ 
&& + \tfrac{118163398}{165375} \pi y^{13/2} + y^7 \Bigl(\tfrac{52369829422440012073}{990186120000000} -  \tfrac{4176344893416403}{990904320000} \pi^2 + \tfrac{351206984461}{6039797760} \pi^4 -  \tfrac{4143716714678}{245581875} \gamma + \tfrac{1753088}{1575} \gamma^2 \nonumber \\ 
&& -  \tfrac{6124042466966}{245581875} \log 2 + \tfrac{7012352}{1575} \gamma \log 2 + \tfrac{7012352}{1575} \log^{2} 2 -  \tfrac{214350489}{30800} \log 3 + \tfrac{9765625}{14256} \log 5 -  \tfrac{2071858357339}{245581875} \log y \nonumber \\ 
&& + \tfrac{1753088}{1575} \gamma \log y + \tfrac{3506176}{1575} \log 2 \log y + \tfrac{438272}{1575} \log^{2} y -  \tfrac{32768}{15} \zeta(3)\Bigr) + \tfrac{169822838237}{245581875} \pi y^{15/2} \nonumber \\ 
&& + y^8 \Bigl(\tfrac{1234405086766291756855079}{10812832430400000000} -  \tfrac{20516582870304319}{9754214400000} \pi^2 -  \tfrac{4004468043930067}{11596411699200} \pi^4 + \tfrac{6403209826927357}{335219259375} \gamma -  \tfrac{819289024}{165375} \gamma^2 \nonumber \\ 
&& + \tfrac{18668500151420029}{335219259375} \log 2 -  \tfrac{4048635776}{165375} \gamma \log 2 -  \tfrac{4434375616}{165375} \log^{2} 2 -  \tfrac{4137804755289}{196196000} \log 3 + \tfrac{227448}{49} \gamma \log 3 \nonumber \\ 
&& + \tfrac{227448}{49} \log 2 \log 3 + \tfrac{113724}{49} \log^{2} 3 -  \tfrac{8837890625}{1111968} \log 5 + \tfrac{6300230470447357}{670438518750} \log y -  \tfrac{819289024}{165375} \gamma \log y \nonumber \\ 
&& -  \tfrac{2024317888}{165375} \log 2 \log y + \tfrac{113724}{49} \log 3 \log y -  \tfrac{204822256}{165375} \log^{2} y + \tfrac{10678144}{1575} \zeta(3)\Bigr) \nonumber \\ 
&& + \bigl(- \tfrac{1048639996225198903}{58998589650000} \pi -  \tfrac{3506176}{4725} \pi^3 + \tfrac{375160832}{165375} \pi \gamma + \tfrac{750321664}{165375} \pi \log 2 + \tfrac{187580416}{165375} \pi \log y\bigr) y^{17/2} + \mathcal{O}\bigl(y^9\bigr). \label{pn:deltaK3+}
\end{eqnarray}

\begin{table}
\begin{center}
\begin{tabular}{c | c }
\hline\hline
$r_\Omega/M$ & $\hat{\delta}_{K3+}$		\\
\hline
$4$		& $-1.072402291940$\\
$5$		& $-0.952268599881$\\
$6$		& $-1.150905925689$\\
$7$		& $-1.347913915585$\\
$8$		& $-1.511472597166$\\
$9$		& $-1.643850731891$\\
$10$	& $-1.751437199028$\\
$12$	& $-1.913557269058$\\
$14$	& $-2.028682058336$\\
$16$	& $-2.114109122984$\\
$18$	& $-2.179795496907$\\
$20$	& $-2.231771587180$\\
$30$	& $-2.383995972376$\\
$40$	& $-2.457665106706$\\
$50$	& $-2.500976521370$\\
$60$	& $-2.529455493583$\\
$70$	& $-2.549596340465$\\
$80$	& $-2.564588968234$\\
$90$	& $-2.576181641423$\\
$100$	& $-2.585412146067$\\
$500$	& $-2.650685806947$\\
$1000$	& $-2.658693616512$\\
$5000$	& $-2.665074853918$\\
\hline
\end{tabular}
\end{center}
\caption{Sample numerical results for the $\hat{\delta}_{K3+}$ as defined in Eq.~\eqref{eq:hatdeltaK3p}.}\label{table:dK3p}
\end{table}

\subsection{Informing initial data models}
How does a black hole move through and respond to an external environment? This question has been addressed by Manasse \cite{Manasse:1963}, and others \cite{Thorne:1984mz, Alvi:1999cw, Detweiler:2000gt, Detweiler:2005kq, Poisson:2005pi, Ishii:2005xq, JohnsonMcDaniel:2009dq, Taylor:2008xy, Poisson:2014gka}, via the method of matched asymptotic expansions (MAE). In scenarios with two distinct length scales ($M \gg \mu$), one may attempt to match `inner' and `outer' expansions across a suitable `buffer' zone ($\mu \ll r \ll M$) \cite{Pound:2010pj}. Indeed, this method was applied to derive the equations of motion underpinning the self-force approach \cite{Mino:1996nk}. Recently, much work has gone into improving initial data for simulations of binary black hole inspirals using MAEs \cite{Yunes:2005nn, Yunes:2006iw, JohnsonMcDaniel:2009dq, Gallouin:2012kb,Mundim:2013vca, Zlochower:2015baa}. 

In a standard approach \cite{Thorne:1984mz, Detweiler:2005kq, Poisson:2005pi}, the black hole is tidally distorted by `external multipole moments': spatial, symmetric, tracefree (STF) tensors $\mathbb{E}_{ij}$, $\mathbb{B}_{ij}$, $\mathbb{E}_{ijk}$, $\mathbb{B}_{ijk}$, etc., related to the Riemann tensor evaluated on the worldline in the \emph{regular} perturbed spacetime. These STF tensors are essentially equivalent to our tetrad-resolved quantities; for example, Detweiler's \cite{Detweiler:2005kq} STF moments are given by $\mathbb{E}_{ij} = \EE_{ij}$, $\mathbb{B}_{ij} = \BB_{ij}$, $\mathbb{E}_{ijk} = \EE_{(ijk)}$ and $\mathbb{B}_{ijk} = \tfrac{3}{4}\BB_{(ijk)}$, with the subtlety of the interchange of spatial indices $2 \leftrightarrow 3$. % NB Swapping the legs over also accounts for the apparent difference of sign in the definitions of the magnetic quantities. 

%\subsubsection{Consistency of Post-Newtonian expansions}
Johnson-McDaniel \emph{et al.} \cite{JohnsonMcDaniel:2009dq} have applied the MAE method to `stitch' two tidally-perturbed Schwarzschild black holes into an external PN metric. Implicit in Eqs.~(B1a)--(B1d) of Ref.~\cite{JohnsonMcDaniel:2009dq} is a PN expansion of (conservative) quadrupolar and octupolar tidal quantities. Restricting to $\mathcal{O}(\mu)$, in our notation Eqs.~(B1a)--(B1d) of \cite{JohnsonMcDaniel:2009dq} imply
\begin{eqnarray}
M^3 \EE_{(111)} &=&  \phantom{-} 6y^4 + 3y^5 + \frac{\mu}{M}\left(-8y^4 + 8 y^5\right) + \mathcal{O}(y^6, \mu^2)  \\
M^3 \EE_{(122)} &=& -3y^4 - 4y^5 + \frac{\mu}{M}\left(4y^4 - \frac{7}{3} y^5\right) + \mathcal{O}(y^6, \mu^2)  \\
M^3 \EE_{(133)} &=& -3y^4 + \phantom{1}y^5 + \frac{\mu}{M}\left(4y^4 - \frac{17}{3} y^5\right) + \mathcal{O}(y^6, \mu^2) \\
M^3 \BB_{(211)} &=&  \phantom{-} 8y^{9/2} + \frac{\mu}{M}\left(-8 y^{9/2}\right)  + \mathcal{O}(y^{11/2}, \mu^2)  \\
M^3 \BB_{(222)} &=& -6 y^{9/2} + \frac{\mu}{M}\left(6 y^{9/2}\right) + \mathcal{O}(y^{11/2}, \mu^2)  \\
M^3 \BB_{(233)} &=& -2 y^{9/2} + \frac{\mu}{M}\left(2 y^{9/2}\right) + \mathcal{O}(y^{11/2}, \mu^2) 
\end{eqnarray}
Note that here the $\mathcal{O}(\mu^0)$ terms are leading-order terms in the Taylor expansion of the `background' Schwarzschild results, and the $\mathcal{O}(\mu^1)$ terms are consistent with the leading terms of our PN series in Sec.~\ref{subsec:pn}. This reassuring consistency suggests that our $\mathcal{O}(\mu/M)$ results may indeed be used to help improve initial data for large mass-ratio binaries in the latter stages of inspiral.

\section{Discussion and conclusion}\label{sec:conclusions}

In the preceding sections we have pursued the line of enquiry of Refs.~\cite{Detweiler:2008ft, Dolan:2013roa, Dolan:2014pja, Bini:2014zxa}, concerned with identifying and calculating $\mathcal{O}(\mu)$ invariants for circular orbits, onwards into the octupolar sector. We identified 7 independent degrees of freedom in the octupolar sector, given by the (symmetrized) components of the derivative of the Riemann tensor as decomposed in the electric-quadrupole basis. A complete set of octupolar invariants for circular orbits is given by, e.g., $\Delta\EE_{(111)}, \Delta\EE_{(122)}, \Delta \BB_{(211)}, \Delta \BB_{(222)}$, $\Delta\EE_{(311)}$, $\Delta\EE_{(322)}$, $\Delta\BB_{(123)}$. Here, the first four are conservative and the latter three are dissipative in character. The remaining symmetrized components $\Delta \EE_{(133)}$, $\Delta \BB_{(233)}$ (conservative) and $\Delta \EE_{(333)}$ (dissipative) follow from trace conditions. All additional octupolar components, $\Delta \EE_{ij0}$, $\Delta \BB_{ij0}$, $\Delta \EE_{i[jk]}$ and $\Delta \BB_{i[jk]}$, may be written in terms of the previous-known quadrupolar tidal invariants $\Delta \EE_{11}$, $\Delta \EE_{22}$, $\Delta \BB_{12}$, $\Delta \BB_{23}$ \cite{Dolan:2014pja}, the spin-precession invariant $\Delta \psi$ \cite{Dolan:2013roa} and the redshift invariant $\Delta U$ \cite{Detweiler:2008ft}. Accurate results for the latter quantities are provided in Tables I \& III of Ref.~\cite{Dolan:2014pja} and PN series are given in Ref.~\cite{Kavanagh:2015lva}. In passing, we should note a relationship which was overlooked in Ref.~\cite{Dolan:2014pja}: $\Delta \BB_{23} = -\overline{\BB}_{12} \Delta \chi$, where $\Delta \chi$ is the dissipative invariant of Table I in Ref.~\cite{Dolan:2014pja}. Also, we should recall that the dissipative component of the self-force, $F_t$ and $F_\phi$, are also invariants. Taken together, we believe we have now arrived at a complete characterization of all circular-orbit invariants in the regular perturbed spacetime through $\mathcal{O}(\mu)$, up to third-derivative order. 

Highly-accurate numerical results for all the octupolar invariants we identify are given in Tables \ref{table:cons_electric}--\ref{table:dK3p}. Our numerical calculation is performed using Mathematica and is made within the Regge-Wheeler gauge as described in Sec.~\ref{sec:computational_approaches}. In addition, as a cross-check on our results, we performed the same calculation in the Lorenz gauge using a Mathematica re-implementation of Ref.~\cite{Akcay:2010} -- see Fig.~\ref{fig:RW_vs_Lorenz} for an example of the excellent agreement we find between the two calculations. To complement our numerical results, we also calculate high-order post-Newtonian expansions for all the invariants. Our technique is briefly described in Sec.~\ref{sec:PN_method} with the full details given in Ref.~\cite{Kavanagh:2015lva}. The lower-order PN expansions are given in Sec.~\ref{subsec:pn} with the higher-order terms available online \cite{online}. In Sec.~\ref{sec:applications} we explored two possible applications for the octupolar invariants.

We can envisage several ways this work could be extended. First, the high-order post-Newtonian results and  the strong-field numerical data could be combined to produce \emph{global} semi-analytic fits for the various invariants. Here, knowledge of the behaviour at the light-ring (Sec.~\ref{sec:light-ring_results}) should prove useful. Similar fits for other invariants have already been applied to EOB models \cite{Akcay:2012ea,Bini:2014ica,Bini:2015kja} and freshly-calibrated EOB models have been successfully compared against numerical relativity simulations \cite{Bernuzzi:2014owa}. Second, we note that in Sec.~\ref{sec:formulation} we have, in fact, derived the form of the octupolar invariants for circular, equatorial orbits in a \emph{rotating} black hole spacetime. Looking ahead, practical calculations on Kerr spacetime are needed. The redshift invariant has already been calculated for circular, equatorial orbits about a Kerr black hole \cite{Shah:2012gu,Isoyama:2014mja}. It seems a natural extension to extend other invariants, such as the ones we describe here, to the rotating scenario. We believe this should be pursued with both numerical and high-order post-Newtonian treatments. Third, a further natural extension is to consider invariants for non-circular orbits. This was recently explored by Akcay \etal~\cite{Akcay:2015pza} for the redshift invariant and we expect the calculation for other invariants to follow in time. Fourth, looking further into the future, invariants at second order in the mass ratio could be calculated. The necessary regularization procedure is now known \cite{Pound:2012prl,Gralla:2012prd,Detweiler:2012prd} and the framework for making practical calculations is beginning to emerge \cite{Pound-Miller:2014,Warburton-Wardell:2013}. As with previous calculations, initial work will focus on the redshift invariant \cite{Pound:2014prd} but the calculation of other invariants will surely follow.

%\NW{Sam: fold this in somewhere?} Mention that first-order spin-dependent corrections arise at relative 1.5PN order: Bini \& Geralico \cite{Bini:2015kja}

\acknowledgements

This material is based upon work supported by the National Science Foundation
under Grant Number 1417132. B.W. was supported by the John Templeton Foundation
New Frontiers Program under Grant No.~37426 (University of Chicago) -
FP050136-B (Cornell University) and by the Irish Research Council, which is
funded under the National Development Plan for Ireland. N.W. gratefully
acknowledges support from a Marie Curie International Outgoing Fellowship
(PIOF-GA-2012-627781) and the Irish Research Council, which is funded under the
National Development Plan for Ireland. S.D.~acknowledges support from the 
Lancaster-Manchester-Sheffield Consortium for Fundamental Physics under STFC grant ST/L000520/1. P.N. and A.C.O. acknowledge support
from Science Foundation Ireland under Grant No. 10/RFP/PHY2847. C.K. is funded under the Programme for Research
in Third Level Institutions (PRTLI) Cycle 5 and cofunded
under the European Regional Development Fund.

\appendix

\section{Gauge invariants in Schwarzschild coordinates}
\label{apdx:invariants_in_Schw_coords}

In this Appendix we give explicit expressions for the perturbations to the
octupolar invariants (as defined in Sec.~\ref{sec:pert-octupole}) for the case
of a circular orbit in Schwarzschild spacetime. Our expressions are written in
terms of the components of $h_{ab}$ and its partial derivatives in
Schwarzschild coordinates, and are given by
\begin{eqnarray}
  \Delta \EE_{(111)} &=& \tfrac{h_{tt} M (8 M - 3 r_0) (3 M - 2 r_0)}{r_0^{5/2} (r_0 - 3 M)^2 (r_0 - 2 M)^{3/2}} + \tfrac{h_{\phi \phi ,r}M^2 (13 M - 6 r_0)}{2 (3 M -  r_0) r_0^{11/2} (r_0 - 2 M)^{1/2}} -  \tfrac{2 h_{r\phi ,\phi r}M (r_0 - 2 M)^{1/2}}{r_0^{9/2}} -  \tfrac{2 h_{tr,\phi r} M^{1/2} (r_0 - 2 M)^{1/2}}{r_0^3} \nonumber \\
&& + \tfrac{h_{t\phi ,rr}M^{1/2} (5 M - 2 r_0) (r_0 - 2 M)^{1/2}}{(3 M -  r_0) r_0^3} -  \tfrac{h_{tt,rr} M (r_0 - 2 M)^{1/2}}{(6 M - 2 r_0) r_0^{3/2}} -  \tfrac{2 h_{r\phi ,\phi }M (6 M -  r_0) (r_0 - 2 M)^{1/2}}{r_0^{11/2} (r_0 - 3 M)} \nonumber \\
&& -  \tfrac{h_{\phi \phi ,rr}M (11 M - 4 r_0) (r_0 - 2 M)^{1/2}}{2 r_0^{9/2} (r_0 - 3 M)} + \tfrac{h_{t\phi ,rrr}M^{1/2} (r_0 - 2 M)^{3/2}}{(3 M -  r_0) r_0^2} -  \tfrac{h_{rr,r} M (6 M - 5 r_0) (r_0 - 2 M)^{3/2}}{2 r_0^{9/2} (r_0 - 3 M)} -  \tfrac{h_{\phi \phi ,rrr}M (r_0 - 2 M)^{3/2}}{2 r_0^{7/2} (r_0 - 3 M)} \nonumber \\
&& -  \tfrac{h_{tt,rrr} (r_0 - 2 M)^{3/2}}{2 r_0^{1/2} (r_0 - 3 M)} + \tfrac{6 h_{t\phi }M^{3/2} (r_0 - 2 M)^{1/2} (4 r_0 - 9 M)}{r_0^5 (r_0 - 3 M)^2} + \tfrac{h_{tt,r} (-47 M^2 + 42 M r_0 - 8 r_0^2)}{2 (3 M -  r_0) r_0^{5/2} (r_0 - 2 M)^{1/2}} + \tfrac{h_{t\phi ,r}M^{1/2} (-35 M^2 + 30 M r_0 - 6 r_0^2)}{(3 M -  r_0) r_0^4 (r_0 - 2 M)^{1/2}} \nonumber \\
&& + \tfrac{2 h_{tr,\phi } M^{1/2} (12 M^2 - 8 M r_0 + r_0^2)}{r_0^4 (r_0 - 3 M) (r_0 - 2 M)^{1/2}} + \tfrac{h_{\phi \phi }M (r_0 - 2 M)^{1/2} (27 M^2 - 18 M r_0 + 4 r_0^2)}{r_0^{13/2} (r_0 - 3 M)^2} -  \tfrac{h_{rr} M (r_0 - 2 M)^{1/2} (66 M^2 - 73 M r_0 + 18 r_0^2)}{2 r_0^{11/2} (r_0 - 3 M)},
\end{eqnarray}
\begin{eqnarray}
  \Delta \EE_{(122)} &=& - \tfrac{2 h_{\theta \phi ,\phi \theta }M}{3 r_0^{11/2} (r_0 - 2 M)^{1/2}} -  \tfrac{2 h_{t\theta ,\phi \theta }M^{1/2}}{3 r_0^4 (r_0 - 2 M)^{1/2}} + \tfrac{h_{tt,r} (14 M - 9 r_0)}{6 r_0^{5/2} (r_0 - 2 M)^{1/2}} + \tfrac{h_{tt} M (20 M - 9 r_0)}{3 r_0^{5/2} (r_0 - 3 M)^2 (r_0 - 2 M)^{1/2}} \nonumber \\
&& -  \tfrac{2 h_{t\phi ,\theta \theta }M^{1/2} (7 M - 4 r_0)}{3 r_0^4 (r_0 - 3 M) (r_0 - 2 M)^{1/2}} -  \tfrac{5 h_{\phi \phi ,\theta \theta }M (r_0 - 2 M)^{1/2}}{(9 M - 3 r_0) r_0^{11/2}} + \tfrac{h_{\phi \phi ,\theta r\theta }M (r_0 - 2 M)^{1/2}}{(18 M - 6 r_0) r_0^{9/2}} + \tfrac{h_{\phi \phi ,\theta \theta r}M (r_0 - 2 M)^{1/2}}{(9 M - 3 r_0) r_0^{9/2}} \nonumber \\
&& + \tfrac{h_{tt,\theta r\theta } (r_0 - 2 M)^{1/2}}{(18 M - 6 r_0) r_0^{3/2}} + \tfrac{h_{tt,\theta \theta r} (r_0 - 2 M)^{1/2}}{(9 M - 3 r_0) r_0^{3/2}} -  \tfrac{h_{\theta \theta }M (2 M - 5 r_0) (r_0 - 2 M)^{1/2}}{r_0^{13/2} (r_0 - 3 M)} + \tfrac{h_{\theta \theta ,r}M (6 M - 11 r_0) (r_0 - 2 M)^{1/2}}{6 r_0^{11/2} (r_0 - 3 M)} \nonumber \\
&& -  \tfrac{h_{r\theta ,\theta }M (6 M - 5 r_0) (r_0 - 2 M)^{1/2}}{3 r_0^{11/2} (r_0 - 3 M)} + \tfrac{2 h_{r\phi ,\phi }M (6 M -  r_0) (r_0 - 2 M)^{1/2}}{3 r_0^{11/2} (r_0 - 3 M)} + \tfrac{2 h_{tr,\phi } M^{1/2} (6 M -  r_0) (r_0 - 2 M)^{1/2}}{3 r_0^4 (r_0 - 3 M)} \nonumber \\
&& + \tfrac{h_{t\phi ,rr}M^{1/2} (r_0 - 2 M)^{3/2}}{(3 M -  r_0) r_0^3} -  \tfrac{h_{\phi \phi ,rr}M (r_0 - 2 M)^{3/2}}{2 r_0^{9/2} (r_0 - 3 M)} -  \tfrac{h_{tt,rr} (r_0 - 2 M)^{3/2}}{2 r_0^{3/2} (r_0 - 3 M)} -  \tfrac{h_{\phi \phi }M^2 (r_0 - 2 M)^{1/2} (18 M + r_0)}{3 r_0^{13/2} (r_0 - 3 M)^2} \nonumber \\
&& + \tfrac{h_{tt,\theta \theta } (-4 M + 3 r_0)}{3 r_0^{5/2} (r_0 - 3 M) (r_0 - 2 M)^{1/2}} -  \tfrac{h_{\phi \phi ,r}M (6 M^2 + 9 M r_0 - 5 r_0^2)}{6 (3 M -  r_0) r_0^{11/2} (r_0 - 2 M)^{1/2}} + \tfrac{h_{t\phi ,r}M^{1/2} (18 M^2 - 25 M r_0 + 7 r_0^2)}{3 (3 M -  r_0) r_0^4 (r_0 - 2 M)^{1/2}} \nonumber \\
&& -  \tfrac{2 h_{t\phi }M^{3/2} (36 M^2 - 38 M r_0 + 11 r_0^2)}{3 r_0^5 (r_0 - 3 M)^2 (r_0 - 2 M)^{1/2}} + \tfrac{h_{rr} M (r_0 - 2 M)^{1/2} (36 M^2 - 56 M r_0 + 19 r_0^2)}{6 r_0^{11/2} (r_0 - 3 M)} + \tfrac{h_{t\phi ,\theta r\theta }M^{1/2} (r_0 - 2 M)^{1/2}}{9 M r_0^3 - 3 r_0^4} \nonumber \\
&& + \tfrac{2 h_{t\phi ,\theta \theta r}M^{1/2} (r_0 - 2 M)^{1/2}}{9 M r_0^3 - 3 r_0^4},
\end{eqnarray}
\begin{eqnarray}
  \Delta \EE_{(133)} &=& - \tfrac{h_{\phi \phi ,\phi \phi }M (M -  r_0)}{r_0^{11/2} (r_0 - 2 M)^{3/2}} -  \tfrac{2 h_{t\phi ,\phi \phi }M^{1/2} (M -  r_0)}{r_0^4 (r_0 - 2 M)^{3/2}} -  \tfrac{h_{\phi \phi ,\phi r\phi }M}{6 r_0^{9/2} (r_0 - 2 M)^{1/2}} -  \tfrac{h_{\phi \phi ,\phi \phi r}M}{3 r_0^{9/2} (r_0 - 2 M)^{1/2}} \nonumber \\
&& -  \tfrac{h_{t\phi ,\phi r\phi }M^{1/2}}{3 r_0^3 (r_0 - 2 M)^{1/2}} -  \tfrac{2 h_{t\phi ,\phi \phi r}M^{1/2}}{3 r_0^3 (r_0 - 2 M)^{1/2}} + \tfrac{h_{tt,r} (23 M - 9 r_0)}{6 r_0^{5/2} (r_0 - 2 M)^{1/2}} -  \tfrac{h_{tt,\phi r\phi }}{6 r_0^{3/2} (r_0 - 2 M)^{1/2}} \nonumber \\
&& -  \tfrac{h_{tt,\phi \phi r}}{3 r_0^{3/2} (r_0 - 2 M)^{1/2}} -  \tfrac{2 h_{r\phi ,\phi r}M (r_0 - 2 M)^{1/2}}{3 r_0^{9/2}} -  \tfrac{2 h_{tr,\phi r} M^{1/2} (r_0 - 2 M)^{1/2}}{3 r_0^3} + \tfrac{2 h_{t\phi }M^{3/2} (M -  r_0) (r_0 - 2 M)^{1/2}}{r_0^5 (r_0 - 3 M)^2} \nonumber \\
&& -  \tfrac{h_{r\phi ,\phi }M (30 M - 11 r_0) (r_0 - 2 M)^{1/2}}{3 r_0^{11/2} (r_0 - 3 M)} -  \tfrac{h_{\phi \phi ,rr}M (5 M -  r_0) (r_0 - 2 M)^{1/2}}{6 r_0^{9/2} (r_0 - 3 M)} + \tfrac{h_{tt,rr} (7 M - 3 r_0) (r_0 - 2 M)^{1/2}}{6 r_0^{3/2} (r_0 - 3 M)} \nonumber \\
&& -  \tfrac{2 h_{rr,r} M (r_0 - 2 M)^{3/2}}{3 r_0^{9/2}} + \tfrac{h_{tt,\phi \phi } (r_0 - M)}{r_0^{5/2} (r_0 - 2 M)^{3/2}} + \tfrac{h_{t\phi ,rr}M^{1/2} (r_0 - 2 M)^{1/2} (r_0 - M)}{3 (3 M -  r_0) r_0^3} + \tfrac{h_{tr,\phi } M^{1/2} (6 M^2 - 7 M r_0 + 2 r_0^2)}{3 r_0^4 (r_0 - 3 M) (r_0 - 2 M)^{1/2}} \nonumber \\
&& + \tfrac{h_{t\phi ,r}M^{1/2} (63 M^2 - 53 M r_0 + 11 r_0^2)}{3 (3 M -  r_0) r_0^4 (r_0 - 2 M)^{1/2}} + \tfrac{h_{\phi \phi ,r}M (147 M^2 - 125 M r_0 + 26 r_0^2)}{6 (3 M -  r_0) r_0^{11/2} (r_0 - 2 M)^{1/2}} + \tfrac{h_{rr} M (r_0 - 2 M)^{1/2} (150 M^2 - 125 M r_0 + 27 r_0^2)}{6 r_0^{11/2} (r_0 - 3 M)} \nonumber \\
&& -  \tfrac{h_{tt} M (-66 M^3 + 111 M^2 r_0 - 56 M r_0^2 + 9 r_0^3)}{3 r_0^{7/2} (r_0 - 3 M)^2 (r_0 - 2 M)^{3/2}} + \tfrac{h_{\phi \phi }M (-222 M^3 + 267 M^2 r_0 - 104 M r_0^2 + 13 r_0^3)}{3 r_0^{13/2} (r_0 - 3 M)^2 (r_0 - 2 M)^{1/2}},
\end{eqnarray}
\begin{eqnarray}
  \Delta \EE_{(113)} &=& \tfrac{h_{\phi \phi ,\phi r}M (3 M -  r_0)}{3 r_0^5 (r_0 - 3 M)^{1/2} (r_0 - 2 M)^{1/2}} + \tfrac{4 h_{t\phi ,\phi r}M^{1/2} (3 M -  r_0)}{3 r_0^{7/2} (r_0 - 3 M)^{1/2} (r_0 - 2 M)^{1/2}} + \tfrac{h_{tt,\phi r} (3 M -  r_0)}{r_0^2 (r_0 - 3 M)^{1/2} (r_0 - 2 M)^{1/2}} \nonumber \\
&& -  \tfrac{2 h_{r\phi ,\phi \phi }M (r_0 - 3 M)^{1/2}}{3 r_0^5 (r_0 - 2 M)^{1/2}} -  \tfrac{2 h_{tr,\phi \phi } M^{1/2} (r_0 - 3 M)^{1/2}}{3 r_0^{7/2} (r_0 - 2 M)^{1/2}} -  \tfrac{h_{\phi \phi ,\phi rr}M (r_0 - 2 M)^{1/2}}{2 r_0^4 (r_0 - 3 M)^{1/2}} -  \tfrac{h_{t\phi ,\phi rr}M^{1/2} (r_0 - 2 M)^{1/2}}{r_0^{5/2} (r_0 - 3 M)^{1/2}} \nonumber \\
&& -  \tfrac{h_{tt,\phi rr} (r_0 - 2 M)^{1/2}}{2 r_0 (r_0 - 3 M)^{1/2}} + \tfrac{h_{rr,\phi } M (r_0 - 2 M)^{1/2} (-12 M + 7 r_0)}{3 r_0^5 (r_0 - 3 M)^{1/2}} + \tfrac{h_{\phi \phi ,\phi }M (-15 M^3 + 36 M^2 r_0 - 17 M r_0^2 + 2 r_0^3)}{6 r_0^6 (r_0 - 3 M)^{3/2} (r_0 - 2 M)^{3/2}} \nonumber \\
&& + \tfrac{h_{tt,\phi } (-17 M^3 + 32 M^2 r_0 - 15 M r_0^2 + 2 r_0^3)}{2 r_0^3 (r_0 - 3 M)^{3/2} (r_0 - 2 M)^{3/2}} + \tfrac{h_{t\phi ,\phi }M^{1/2} (-87 M^3 + 120 M^2 r_0 - 49 M r_0^2 + 6 r_0^3)}{3 r_0^{9/2} (r_0 - 3 M)^{3/2} (r_0 - 2 M)^{3/2}},
\end{eqnarray}
\begin{eqnarray}
  \Delta \EE_{(223)} &=& - \tfrac{h_{\phi \phi ,\phi \theta \theta }M}{2 r_0^5 (r_0 - 3 M)^{1/2} (r_0 - 2 M)^{1/2}} + \tfrac{2 h_{\phi \phi ,\phi r}M^2}{3 r_0^5 (r_0 - 3 M)^{1/2} (r_0 - 2 M)^{1/2}} -  \tfrac{4 h_{t\theta ,\theta }M^{3/2}}{3 r_0^{9/2} (r_0 - 3 M)^{1/2} (r_0 - 2 M)^{1/2}} -  \tfrac{h_{t\phi ,\phi \theta \theta }M^{1/2}}{r_0^{7/2} (r_0 - 3 M)^{1/2} (r_0 - 2 M)^{1/2}} \nonumber \\
&& + \tfrac{4 h_{t\phi ,\phi r}M^{3/2}}{3 r_0^{7/2} (r_0 - 3 M)^{1/2} (r_0 - 2 M)^{1/2}} -  \tfrac{h_{tt,\phi \theta \theta }}{2 r_0^2 (r_0 - 3 M)^{1/2} (r_0 - 2 M)^{1/2}} + \tfrac{2 h_{tt,\phi r} M}{3 r_0^2 (r_0 - 3 M)^{1/2} (r_0 - 2 M)^{1/2}} -  \tfrac{4 h_{\theta \phi ,\theta }M (r_0 - 2 M)^{1/2}}{3 r_0^6 (r_0 - 3 M)^{1/2}} \nonumber \\
&& + \tfrac{2 h_{\theta \phi ,\theta r}M (r_0 - 2 M)^{1/2}}{3 r_0^5 (r_0 - 3 M)^{1/2}} -  \tfrac{2 h_{r\phi ,\theta \theta }M (r_0 - 2 M)^{1/2}}{3 r_0^5 (r_0 - 3 M)^{1/2}} -  \tfrac{4 h_{r\phi }M (r_0 - 2 M)^{1/2}}{3 r_0^5 (r_0 - 3 M)^{1/2}} + \tfrac{2 h_{t\theta ,\theta r}M^{1/2} (r_0 - 2 M)^{1/2}}{3 r_0^{7/2} (r_0 - 3 M)^{1/2}} \nonumber \\
&& -  \tfrac{2 h_{tr,\theta \theta } M^{1/2} (r_0 - 2 M)^{1/2}}{3 r_0^{7/2} (r_0 - 3 M)^{1/2}} -  \tfrac{h_{\theta \theta ,\phi }M (6 M + r_0)}{3 r_0^6 (r_0 - 3 M)^{1/2} (r_0 - 2 M)^{1/2}} + \tfrac{h_{tt,\phi } (-6 M^3 - 11 M^2 r_0 + 14 M r_0^2 - 3 r_0^3)}{6 r_0^3 (r_0 - 3 M)^{3/2} (r_0 - 2 M)^{3/2}} \nonumber \\
&& -  \tfrac{h_{\phi \phi ,\phi }M (6 M^3 + 11 M^2 r_0 - 14 M r_0^2 + 3 r_0^3)}{6 r_0^6 (r_0 - 3 M)^{3/2} (r_0 - 2 M)^{3/2}} -  \tfrac{h_{t\phi ,\phi }M^{1/2} (6 M^3 + 11 M^2 r_0 - 14 M r_0^2 + 3 r_0^3)}{3 r_0^{9/2} (r_0 - 3 M)^{3/2} (r_0 - 2 M)^{3/2}},
\end{eqnarray}
\begin{eqnarray}
  \Delta \EE_{(333)} &=& - \tfrac{h_{\phi \phi ,\phi \phi \phi }M (r_0 - 3 M)^{1/2}}{2 r_0^5 (r_0 - 2 M)^{3/2}} -  \tfrac{h_{t\phi ,\phi \phi \phi }M^{1/2} (r_0 - 3 M)^{1/2}}{r_0^{7/2} (r_0 - 2 M)^{3/2}} -  \tfrac{h_{tt,\phi \phi \phi } (r_0 - 3 M)^{1/2}}{2 r_0^2 (r_0 - 2 M)^{3/2}} -  \tfrac{h_{tt,\phi } (r_0 - 3 M)^{3/2}}{2 r_0^3 (r_0 - 2 M)^{3/2}} \nonumber \\
&& + \tfrac{h_{\phi \phi ,\phi }M (7 M - 5 r_0) (3 M -  r_0)}{2 (2 M -  r_0) r_0^6 (r_0 - 3 M)^{1/2} (r_0 - 2 M)^{1/2}} + \tfrac{2 h_{\phi \phi ,\phi r}M (r_0 - 3 M)^{1/2}}{r_0^5 (r_0 - 2 M)^{1/2}} -  \tfrac{2 h_{r\phi ,\phi \phi }M (r_0 - 3 M)^{1/2}}{r_0^5 (r_0 - 2 M)^{1/2}} + \tfrac{2 h_{t\phi ,\phi r}M^{1/2} (r_0 - 3 M)^{1/2}}{r_0^{7/2} (r_0 - 2 M)^{1/2}} \nonumber \\
&& -  \tfrac{2 h_{tr,\phi \phi } M^{1/2} (r_0 - 3 M)^{1/2}}{r_0^{7/2} (r_0 - 2 M)^{1/2}} -  \tfrac{2 h_{rr,\phi } M (r_0 - 3 M)^{1/2} (r_0 - 2 M)^{1/2}}{r_0^5} + \tfrac{h_{t\phi ,\phi }M^{1/2} (3 M -  r_0) (M + r_0)}{r_0^{9/2} (r_0 - 3 M)^{1/2} (r_0 - 2 M)^{3/2}},
\end{eqnarray}
\begin{eqnarray}
  \Delta \BB_{(123)} &=& \tfrac{h_{tr,\theta \theta } (7 M - 2 r_0) (2 M -  r_0)}{3 r_0^{7/2} (r_0 - 3 M)^{3/2}} + \tfrac{h_{\theta \phi ,\theta }M^{3/2}}{3 r_0^6 (r_0 - 3 M)^{1/2}} -  \tfrac{h_{\theta \phi ,\theta r}M^{3/2}}{6 r_0^5 (r_0 - 3 M)^{1/2}} + \tfrac{h_{r\phi ,\theta \theta }M^{3/2}}{6 r_0^5 (r_0 - 3 M)^{1/2}} -  \tfrac{h_{\theta \phi ,\phi \phi \theta }M^{3/2}}{6 (2 M -  r_0) r_0^5 (r_0 - 3 M)^{1/2}} \nonumber \\
&& -  \tfrac{h_{\theta \theta ,\phi \phi \phi }M^{3/2}}{6 (2 M -  r_0) r_0^5 (r_0 - 3 M)^{1/2}} + \tfrac{h_{rr,\phi \theta \theta } M^{1/2} (2 M -  r_0)}{6 r_0^4 (r_0 - 3 M)^{1/2}} -  \tfrac{h_{t\phi ,\phi \theta \theta }M}{(12 M - 6 r_0) r_0^{7/2} (r_0 - 3 M)^{1/2}} -  \tfrac{h_{t\theta ,\phi \phi \theta }M}{(4 M - 2 r_0) r_0^{7/2} (r_0 - 3 M)^{1/2}} \nonumber \\
&& + \tfrac{h_{tr,\theta \theta r} (2 M -  r_0)}{3 r_0^{5/2} (r_0 - 3 M)^{1/2}} -  \tfrac{h_{tt,\phi \theta \theta } M^{1/2}}{3 (2 M -  r_0) r_0^2 (r_0 - 3 M)^{1/2}} -  \tfrac{h_{t\phi,\theta\theta}}{6 r_0^{3/2} (r_0 - 3 M)^{1/2}} -  \tfrac{h_{tt,\phi rr} M^{1/2}}{3 r_0 (r_0 - 3 M)^{1/2}} + \tfrac{h_{tr,\phi \phi r} (r_0 - 4 M)}{6 r_0^{5/2} (r_0 - 3 M)^{1/2}} \nonumber \\
&& + \tfrac{h_{\theta \theta ,\phi rr}M^{1/2} (r_0 - 2 M)}{6 r_0^4 (r_0 - 3 M)^{1/2}} + \tfrac{h_{rr,\phi \phi \phi } M^{1/2} (r_0 - 2 M)}{6 r_0^4 (r_0 - 3 M)^{1/2}} + \tfrac{h_{t\theta ,\theta rr}(r_0 - 2 M)}{3 r_0^{5/2} (r_0 - 3 M)^{1/2}} + \tfrac{h_{rr,\phi r} M^{1/2} (r_0 - 3 M)^{1/2} (r_0 - 2 M)}{6 (3 M -  r_0) r_0^3} \nonumber \\
&& -  \tfrac{h_{r\theta ,\phi \theta }M^{1/2} (-13 M^2 + 17 M r_0 - 4 r_0^2)}{6 r_0^5 (r_0 - 3 M)^{3/2}} + \tfrac{h_{\theta \theta ,\phi r}M^{1/2} (-8 M^2 + 9 M r_0 - 2 r_0^2)}{3 r_0^5 (r_0 - 3 M)^{3/2}} + \tfrac{h_{t\theta ,\theta r}(-8 M^2 + 9 M r_0 - 2 r_0^2)}{3 r_0^{7/2} (r_0 - 3 M)^{3/2}} \nonumber \\
&& + \tfrac{2 h_{t\phi ,\phi }M (M^2 + 4 M r_0 -  r_0^2)}{3 (2 M -  r_0) r_0^{9/2} (r_0 - 3 M)^{3/2}} + \tfrac{h_{\phi \phi ,\phi r}M^{1/2} (32 M^2 - 13 M r_0 + r_0^2)}{6 r_0^5 (r_0 - 3 M)^{3/2}} -  \tfrac{2 h_{t\theta ,\theta }M (20 M^2 - 19 M r_0 + 4 r_0^2)}{3 (2 M -  r_0) r_0^{9/2} (r_0 - 3 M)^{3/2}} \nonumber \\
&& -  \tfrac{h_{r\phi }M (3 M^{3/2} r_0^{3/2} -  M^{1/2} r_0^{5/2})}{3 r_0^{13/2} (r_0 - 3 M)^{3/2}} + \tfrac{2 h_{tt,\phi r} M^{1/2} (r_0 - 3 M)^{1/2}}{6 M r_0^2 - 3 r_0^3} -  \tfrac{h_{r\phi ,\phi \phi }M^{1/2} (28 M^3 - 37 M^2 r_0 + 15 M r_0^2 - 2 r_0^3)}{6 (2 M -  r_0) r_0^5 (r_0 - 3 M)^{3/2}} \nonumber \\
&& + \tfrac{h_{tr,\phi \phi } (52 M^3 - 57 M^2 r_0 + 19 M r_0^2 - 2 r_0^3)}{6 r_0^{7/2} (r_0 - 3 M)^{3/2} (r_0 - 2 M)} -  \tfrac{h_{\phi \phi ,\phi }M^{1/2} (26 M^3 - 36 M^2 r_0 + 11 M r_0^2 -  r_0^3)}{6 (2 M -  r_0) r_0^6 (r_0 - 3 M)^{3/2}} -  \tfrac{h_{t\phi ,\phi r}(-100 M^3 + 91 M^2 r_0 - 25 M r_0^2 + 2 r_0^3)}{6 (2 M -  r_0) r_0^{7/2} (r_0 - 3 M)^{3/2}} \nonumber \\
&& -  \tfrac{h_{\theta \theta ,\phi }M^{1/2} (4 M^3 + 7 M^2 r_0 - 9 M r_0^2 + 2 r_0^3)}{3 (2 M -  r_0) r_0^6 (r_0 - 3 M)^{3/2}} + \tfrac{h_{rr,\phi } (68 M^{7/2} r_0^{1/2} - 45 M^{5/2} r_0^{3/2} + 4 M^{3/2} r_0^{5/2} + M^{1/2} r_0^{7/2})}{6 r_0^{11/2} (r_0 - 3 M)^{3/2}} \nonumber \\
&& + \tfrac{h_{tt,\phi } (-72 M^{7/2} r_0^{1/2} + 109 M^{5/2} r_0^{3/2} - 47 M^{3/2} r_0^{5/2} + 6 M^{1/2} r_0^{7/2})}{6 r_0^{7/2} (r_0 - 3 M)^{3/2} (r_0 - 2 M)^2} + \tfrac{h_{r\phi ,\phi \phi r}M^{1/2} (r_0 - 3 M)^{1/2}}{18 M r_0^3 - 6 r_0^4},
\end{eqnarray}
\begin{eqnarray}
  \Delta \BB_{(211)} &=& - \tfrac{h_{r\theta ,\phi \phi \theta }M^{1/2}}{6 r_0^{9/2}} + \tfrac{h_{rr,\phi \phi } M^{1/2} (4 M - 3 r_0)}{6 r_0^{9/2}} -  \tfrac{h_{tr,\phi \theta \theta } M}{6 (3 M -  r_0) r_0^3} -  \tfrac{2 h_{tt,\theta \theta r} M^{1/2}}{(9 M - 3 r_0) r_0^{3/2}} + \tfrac{h_{t\theta ,\phi \theta r}(r_0 - 8 M)}{6 (3 M -  r_0) r_0^3} + \tfrac{6 h_{\theta \theta }M^{3/2} (2 M -  r_0)}{r_0^{13/2} (r_0 - 3 M)} \nonumber \\
&& -  \tfrac{10 h_{\theta \theta ,r}M^{3/2} (2 M -  r_0)}{3 r_0^{11/2} (r_0 - 3 M)} + \tfrac{16 h_{r\theta ,\theta }M^{3/2} (2 M -  r_0)}{3 r_0^{11/2} (r_0 - 3 M)} + \tfrac{h_{\theta \phi ,\phi \theta r}M^{3/2}}{6 r_0^{9/2} (r_0 - 3 M)} + \tfrac{h_{\theta \theta ,\phi \phi r}M^{3/2}}{3 r_0^{9/2} (r_0 - 3 M)} + \tfrac{h_{\theta \theta ,rr}M^{3/2} (2 M -  r_0)}{3 r_0^{9/2} (r_0 - 3 M)} \nonumber \\
&& -  \tfrac{2 h_{r\theta ,\theta r}M^{3/2} (2 M -  r_0)}{3 r_0^{9/2} (r_0 - 3 M)} + \tfrac{h_{rr,r} M^{3/2} (22 M - 9 r_0) (2 M -  r_0)}{3 r_0^{9/2} (r_0 - 3 M)} + \tfrac{h_{r\phi ,\phi r}M^{1/2} (M -  r_0) (4 M -  r_0)}{6 r_0^{9/2} (r_0 - 3 M)} -  \tfrac{h_{\theta \phi ,\phi \theta }M^{1/2} (M -  r_0)^2}{6 r_0^{11/2} (r_0 - 3 M) (r_0 - 2 M)} \nonumber \\
&& + \tfrac{h_{tt,\theta \theta } M^{1/2} (5 M - 4 r_0)}{3 r_0^{5/2} (r_0 - 3 M) (r_0 - 2 M)} + \tfrac{h_{tr,\phi rr} (r_0 - 2 M)}{3 r_0^2} + \tfrac{h_{tt,rrr} M^{1/2} (r_0 - 2 M)}{(9 M - 3 r_0) r_0^{1/2}} + \tfrac{h_{r\phi ,\phi \theta \theta }M^{1/2} (r_0 - 2 M)}{6 r_0^{9/2} (r_0 - 3 M)} + \tfrac{h_{rr,\theta \theta } M^{3/2} (r_0 - 2 M)}{3 r_0^{9/2} (r_0 - 3 M)} \nonumber \\
&& + \tfrac{h_{rr,\phi \phi r} M^{1/2} (r_0 - 2 M)^2}{3 r_0^{7/2} (r_0 - 3 M)} -  \tfrac{h_{rr,rr} M^{3/2} (r_0 - 2 M)^2}{3 r_0^{7/2} (r_0 - 3 M)} + \tfrac{4 h_{\phi \phi }M^{3/2} (r_0 - 2 M) (r_0 - M)}{r_0^{13/2} (r_0 - 3 M)^2} + \tfrac{h_{tt,rr} M^{1/2} (-3 M + 2 r_0)}{3 r_0^{3/2} (r_0 - 3 M)} \nonumber \\
&& + \tfrac{4 h_{\phi \phi ,r}M^{3/2} (-8 M + 3 r_0)}{3 r_0^{11/2} (r_0 - 3 M)} + \tfrac{h_{t\phi ,\theta \theta }(-9 M^2 + 10 M r_0 - 3 r_0^2)}{6 r_0^4 (r_0 - 3 M) (r_0 - 2 M)} + \tfrac{h_{\theta \theta ,\phi \phi }M^{1/2} (10 M^2 - 9 M r_0 + r_0^2)}{6 r_0^{11/2} (r_0 - 3 M) (r_0 - 2 M)} -  \tfrac{h_{r\phi ,\phi rr}M^{1/2} (2 M^2 - 3 M r_0 + r_0^2)}{3 r_0^{7/2} (r_0 - 3 M)} \nonumber \\
&& + \tfrac{h_{t\theta ,\phi \theta }(31 M^2 - 26 M r_0 + 3 r_0^2)}{6 r_0^4 (r_0 - 3 M) (r_0 - 2 M)} + \tfrac{h_{r\phi ,\phi }M^{1/2} (16 M^2 - 21 M r_0 + 4 r_0^2)}{6 r_0^{11/2} (r_0 - 3 M)} + \tfrac{h_{tt,r} M^{1/2} (70 M^2 - 57 M r_0 + 10 r_0^2)}{3 r_0^{5/2} (r_0 - 3 M) (r_0 - 2 M)} \nonumber \\
&& -  \tfrac{2 h_{rr} M^{3/2} (94 M^2 - 83 M r_0 + 18 r_0^2)}{3 r_0^{11/2} (r_0 - 3 M)} + \tfrac{h_{t\phi ,rrr}(2 M^2 - 3 M r_0 + r_0^2)}{9 M r_0^2 - 3 r_0^3} + \tfrac{h_{tr,\phi } (-112 M^3 + 156 M^2 r_0 - 61 M r_0^2 + 6 r_0^3)}{6 r_0^4 (r_0 - 3 M) (r_0 - 2 M)} \nonumber \\
&& + \tfrac{2 h_{tt} M^{3/2} (-114 M^3 + 189 M^2 r_0 - 103 M r_0^2 + 18 r_0^3)}{3 r_0^{7/2} (r_0 - 3 M)^2 (r_0 - 2 M)^2} + \tfrac{h_{t\phi }M (-120 M^3 + 239 M^2 r_0 - 145 M r_0^2 + 28 r_0^3)}{3 r_0^5 (r_0 - 3 M)^2 (r_0 - 2 M)} + \tfrac{h_{t\phi ,\theta \theta r}(M -  r_0)}{18 M r_0^3 - 6 r_0^4} \nonumber \\
&& + \tfrac{h_{tr,\phi r} (36 M^2 - 29 M r_0 + 5 r_0^2)}{18 M r_0^3 - 6 r_0^4} -  \tfrac{h_{t\phi ,rr}(12 M^2 - 17 M r_0 + 5 r_0^2)}{18 M r_0^3 - 6 r_0^4} + \tfrac{h_{t\phi ,r}(136 M^2 - 63 M r_0 + 6 r_0^2)}{18 M r_0^4 - 6 r_0^5},
\end{eqnarray}
\begin{eqnarray}
  \Delta \BB_{(222)} &=& - \tfrac{h_{\theta \theta ,\phi \phi }M^{1/2}}{2 r_0^{11/2}} + \tfrac{h_{r\theta ,\phi \phi \theta }M^{1/2}}{2 r_0^{9/2}} + \tfrac{h_{r\phi ,\phi \theta \theta }M^{3/2}}{(6 M - 2 r_0) r_0^{9/2}} -  \tfrac{h_{t\phi,\theta\theta}}{(3 M -  r_0) r_0^3} + \tfrac{h_{tt,\theta \theta r} M^{1/2}}{(3 M -  r_0) r_0^{3/2}} + \tfrac{h_{t\theta ,\phi \theta }(r_0 - 5 M)}{(3 M -  r_0) r_0^4} -  \tfrac{8 h_{\theta \theta }M^{3/2} (2 M -  r_0)}{r_0^{13/2} (r_0 - 3 M)} \nonumber \\
&& + \tfrac{h_{\phi \phi ,r}M^{3/2} (8 M - 3 r_0)}{r_0^{11/2} (r_0 - 3 M)} + \tfrac{4 h_{\theta \theta ,r}M^{3/2} (2 M -  r_0)}{r_0^{11/2} (r_0 - 3 M)} -  \tfrac{8 h_{r\theta ,\theta }M^{3/2} (2 M -  r_0)}{r_0^{11/2} (r_0 - 3 M)} + \tfrac{h_{rr} M^{3/2} (20 M - 9 r_0) (2 M -  r_0)}{r_0^{11/2} (r_0 - 3 M)} \nonumber \\
&& + \tfrac{h_{\theta \phi ,\phi \theta r}M^{1/2} (2 M -  r_0)}{2 r_0^{9/2} (r_0 - 3 M)} + \tfrac{h_{rr,\theta \theta } M^{3/2} (2 M -  r_0)}{r_0^{9/2} (r_0 - 3 M)} + \tfrac{h_{tt,r} M^{1/2} (7 M - 2 r_0)}{r_0^{5/2} (r_0 - 3 M)} + \tfrac{h_{tt,rr} M^{1/2} (2 M -  r_0)}{r_0^{3/2} (r_0 - 3 M)} + \tfrac{h_{rr,\phi \phi } M^{1/2} (r_0 - 2 M)}{2 r_0^{9/2}} \nonumber \\
&& + \tfrac{h_{r\phi ,\phi }M^{1/2} (r_0 - 2 M)}{2 r_0^{9/2} (r_0 - 3 M)} -  \tfrac{h_{rr,r} M^{3/2} (r_0 - 2 M)^2}{r_0^{9/2} (r_0 - 3 M)} + \tfrac{h_{\theta \phi ,\phi \theta }M^{1/2} (r_0 - M)}{r_0^{11/2} (r_0 - 3 M)} -  \tfrac{3 h_{\phi \phi }M^{3/2} (r_0 - 2 M) (r_0 - M)}{r_0^{13/2} (r_0 - 3 M)^2} + \tfrac{h_{tt,\theta \theta } M^{1/2} (2 r_0 - 3 M)}{r_0^{5/2} (r_0 - 3 M) (r_0 - 2 M)} \nonumber \\
&& + \tfrac{h_{tt} M^{3/2} (-36 M^2 + 37 M r_0 - 9 r_0^2)}{r_0^{7/2} (r_0 - 3 M)^2 (r_0 - 2 M)} + \tfrac{h_{t\phi }M (-6 M^2 + 7 M r_0 - 3 r_0^2)}{r_0^5 (r_0 - 3 M)^2} -  \tfrac{h_{r\phi ,\phi r}M^{1/2} (2 M^2 - 3 M r_0 + r_0^2)}{2 r_0^{9/2} (r_0 - 3 M)} + \tfrac{h_{t\theta ,\phi \theta r}M}{6 M r_0^3 - 2 r_0^4} \nonumber \\
&& -  \tfrac{h_{t\phi ,\theta \theta r}(M -  r_0)}{6 M r_0^3 - 2 r_0^4} + \tfrac{h_{tr,\phi \theta \theta } (4 M -  r_0)}{6 M r_0^3 - 2 r_0^4} -  \tfrac{h_{tr,\phi r} (10 M^2 - 7 M r_0 + r_0^2)}{6 M r_0^3 - 2 r_0^4} + \tfrac{h_{t\phi ,rr}(2 M^2 - 3 M r_0 + r_0^2)}{6 M r_0^3 - 2 r_0^4} -  \tfrac{h_{t\phi ,r}(36 M^2 - 16 M r_0 + r_0^2)}{6 M r_0^4 - 2 r_0^5} \nonumber \\
&& + \tfrac{h_{tr,\phi } (18 M^2 - 10 M r_0 + r_0^2)}{6 M r_0^4 - 2 r_0^5},
\end{eqnarray}
\begin{eqnarray}
  \Delta \BB_{(233)} &=& \tfrac{h_{\theta \phi ,\phi \theta r}M^{1/2}}{6 r_0^{9/2}} -  \tfrac{h_{r\phi ,\phi \theta \theta }M^{1/2}}{6 r_0^{9/2}} + \tfrac{h_{\theta \theta ,\phi \phi r}M^{1/2}}{3 r_0^{9/2}} -  \tfrac{h_{r\theta ,\phi \phi \theta }M^{1/2}}{3 r_0^{9/2}} + \tfrac{h_{r\phi ,\phi \phi \phi }M^{3/2}}{(6 M - 3 r_0) r_0^{9/2}} + \tfrac{h_{r\phi ,\phi r}M^{5/2}}{(9 M - 3 r_0) r_0^{9/2}} -  \tfrac{h_{rr,\phi \phi } M^{1/2}}{3 r_0^{7/2}} + \tfrac{h_{t\theta,r\theta\phi}}{2 r_0^3} \nonumber \\
&& -  \tfrac{h_{tr,\phi \theta \theta }}{2 r_0^3} + \tfrac{h_{tt,\phi \phi r} M^{1/2}}{(6 M - 3 r_0) r_0^{3/2}} -  \tfrac{2 h_{\theta \theta }M^{3/2} (2 M -  r_0)}{3 r_0^{13/2} (r_0 - 3 M)} + \tfrac{h_{\phi \phi ,\phi \phi }M^{3/2}}{3 r_0^{11/2} (r_0 - 3 M)} + \tfrac{h_{\phi \phi ,r}M^{3/2} (16 M - 7 r_0)}{3 r_0^{11/2} (r_0 - 3 M)} + \tfrac{2 h_{\theta \theta ,r}M^{3/2} (2 M -  r_0)}{3 r_0^{11/2} (r_0 - 3 M)} \nonumber \\
&& + \tfrac{h_{rr} M^{3/2} (22 M - 9 r_0) (2 M -  r_0)}{3 r_0^{11/2} (r_0 - 3 M)} + \tfrac{2 h_{r\theta ,\theta r}M^{3/2} (2 M -  r_0)}{3 r_0^{9/2} (r_0 - 3 M)} + \tfrac{2 h_{tt,r} M^{1/2} (4 M -  r_0)}{3 r_0^{5/2} (r_0 - 3 M)} + \tfrac{h_{tt,rr} M^{1/2} (2 M -  r_0)}{3 r_0^{3/2} (r_0 - 3 M)} \nonumber \\
&& + \tfrac{2 h_{\theta \theta ,\phi \phi }M^{1/2} (M -  r_0)}{3 r_0^{11/2} (r_0 - 2 M)} -  \tfrac{2 h_{t\phi ,\phi \phi }M (5 M - 2 r_0)}{3 r_0^4 (r_0 - 3 M) (r_0 - 2 M)} + \tfrac{h_{tt,\theta \theta } M^{3/2}}{3 r_0^{5/2} (r_0 - 3 M) (r_0 - 2 M)} + \tfrac{h_{\theta \theta ,rr}M^{3/2} (r_0 - 2 M)}{3 r_0^{9/2} (r_0 - 3 M)} + \tfrac{h_{rr,\theta \theta } M^{3/2} (r_0 - 2 M)}{3 r_0^{9/2} (r_0 - 3 M)} \nonumber \\
&& -  \tfrac{2 h_{rr,r} M^{3/2} (r_0 - 2 M)^2}{3 r_0^{9/2} (r_0 - 3 M)} + \tfrac{h_{tr,\phi r} M (2 r_0 - 5 M)}{3 (3 M -  r_0) r_0^3} + \tfrac{h_{\phi \phi }M^{3/2} (r_0 - 2 M) (5 r_0 - 21 M)}{3 r_0^{13/2} (r_0 - 3 M)^2} + \tfrac{h_{t\theta ,\phi \theta }(5 M^2 + 8 M r_0 - 3 r_0^2)}{6 r_0^4 (r_0 - 3 M) (r_0 - 2 M)} \nonumber \\
&& + \tfrac{h_{\theta \phi ,\phi \theta }M^{1/2} (5 M^2 - 4 M r_0 + r_0^2)}{6 r_0^{11/2} (r_0 - 3 M) (r_0 - 2 M)} + \tfrac{h_{r\phi ,\phi }M^{1/2} (-32 M^2 + 11 M r_0 + r_0^2)}{6 r_0^{11/2} (r_0 - 3 M)} + \tfrac{h_{tt,\phi \phi } M^{1/2} (7 M^2 - 8 M r_0 + 2 r_0^2)}{3 r_0^{5/2} (r_0 - 3 M) (r_0 - 2 M)^2} \nonumber \\
&& + \tfrac{h_{t\phi ,\theta \theta }(17 M^2 - 14 M r_0 + 3 r_0^2)}{6 r_0^4 (r_0 - 3 M) (r_0 - 2 M)} -  \tfrac{h_{tt} M^{3/2} (30 M^2 - 35 M r_0 + 9 r_0^2)}{3 r_0^{7/2} (r_0 - 3 M)^2 (r_0 - 2 M)} + \tfrac{h_{tr,\phi } (28 M^3 - 42 M^2 r_0 + 21 M r_0^2 - 3 r_0^3)}{6 r_0^4 (r_0 - 3 M) (r_0 - 2 M)} \nonumber \\
&& + \tfrac{h_{t\phi }M (-12 M^3 + 23 M^2 r_0 - 14 M r_0^2 + 3 r_0^3)}{3 (2 M -  r_0) r_0^5 (r_0 - 3 M)^2} + \tfrac{h_{t\phi ,\phi \phi r}M}{6 M r_0^3 - 3 r_0^4} + \tfrac{h_{tr,\phi \phi \phi } M}{6 M r_0^3 - 3 r_0^4} + \tfrac{h_{t\phi ,rr}M^2}{9 M r_0^3 - 3 r_0^4} -  \tfrac{h_{t\phi ,r}(36 M^2 - 23 M r_0 + 3 r_0^2)}{18 M r_0^4 - 6 r_0^5}.
\end{eqnarray}

\section{Shift to asymptotically flat gauge}
\label{sec:shift-asympt}

In order to compare our results with PN theory it is necessary to work in an
asymptotically flat gauge. In both the Lorenz and Zerilli gauges the
$tt$-component of the metric perturbation does not vanish at spatial infinity
and so we make an $\mathcal{O}(\mu)$ gauge transformation to correct for this
\cite{Sago:2008id}. For both gauges this correction can be made
by adding $h^{NAF}_{ab} = \xi_{a ; b} + \xi_{b ; a}$ where
$\xi{^a} = [- \alpha(t + r_* - r), 0, 0, 0]$ and
$\alpha = \mu/\sqrt{r_0(r_0-3M)}$. Explicitly, this can be achieved by adding an
extra term to the invariants,
$\Delta \EE_{(ijk)} \rightarrow \Delta \EE_{(ijk)} + \delta^\xi \EE$ and
$\Delta \BB_{(ijk)} \rightarrow \Delta \BB_{(ijk)} + \delta^\xi \EE$ where
\begin{subequations}
\begin{eqnarray}
  \delta^\xi \EE_{(111)} &=& \frac{2 \alpha  M \left(-81 M^3+111 M^2 r_0-51 M r_0^2+8 r_0^3\right)}{r_0^{9/2} (r_0-3 M)^2 (r_0-2 M)^{1/2}}, \\
  \delta^\xi \EE_{(122)} &=& \frac{2 \alpha  M \left(54 M^3-109 M^2 r_0+64 M r_0^2-12 r_0^3\right)}{3 r_0^{9/2} (r_0-3 M)^2 (r_0-2 M)^{1/2}}, \\
  \delta^\xi \EE_{(133)} &=& \frac{2 \alpha  M \left(189 M^3-224 M^2 r_0+89 M r_0^2-12 r_0^3\right)}{3 r_0^{9/2} (r_0-3 M)^2 (r_0-2 M)^{1/2}}, \\
  \delta^\xi \EE_{(113)} &=& 0, \\
  \delta^\xi \EE_{(223)} &=& 0, \\
  \delta^\xi \EE_{(333)} &=& 0, \\
  \delta^\xi \BB_{(123)} &=& 0, \\
  \delta^\xi \BB_{(211)} &=& \frac{8 \alpha  M^{3/2} \left(54 M^2-43 M r_0+9 r_0^2\right)}{3 r_0^{9/2} (r_0-3 M)^2}, \\
  \delta^\xi \BB_{(222)} &=& -\frac{2 \alpha  M^{3/2} \left(54 M^2-43 M r_0+9 r_0^2\right)}{r_0^{9/2} (r_0-3 M)^2}, \\
  \delta^\xi \BB_{(233)} &=& -\frac{2 \alpha  M^{3/2} \left(54 M^2-43 M r_0+9 r_0^2\right)}{3 r_0^{9/2} (r_0-3 M)^2}.
\end{eqnarray}
\end{subequations}

\bibliographystyle{apsrev4-1}
\bibliography{references}

\end{document}